\documentclass[reprint,superscriptaddress,amsmath,amssymb,aps, longbibliography,pre]{revtex4-2}

\usepackage{graphicx}
\usepackage{bm}
\usepackage{siunitx}
\usepackage[noend]{algorithmic}
\usepackage{algorithm}
\usepackage[skins]{tcolorbox}
\usepackage{setspace}
\usepackage{blkarray}
\usepackage{cancel}
\usepackage{xcolor}
\usepackage{soul}
\usepackage{systeme}
\usepackage{scalerel} 

\makeatletter
\def\maketag@@@#1{\hbox{\m@th\normalfont\normalsize#1}}
\makeatother

\newcommand{\nn}{\nonumber\\}

\usepackage{hyperref}
\usepackage[capitalize]{cleveref}

\usepackage[utf8]{inputenc}

\usepackage[lining,semibold]{libertine} 
\usepackage{amsthm}
\usepackage[libertine, cmintegrals, bigdelims, vvarbb]{newtxmath}
\usepackage{mathrsfs}
\usepackage{gensymb}
\usepackage{bbm}
\usepackage{dsfont}
\usepackage{pgfplots}
\pgfplotsset{compat=1.7}
\usepgfplotslibrary{colormaps}
\usepackage{kbordermatrix}
\renewcommand{\kbldelim}{(}
\renewcommand{\kbrdelim}{)}
\usepackage{chemformula}
\usepackage[caption=false]{subfig}

\theoremstyle{definition}

\usepackage{scalerel} 

\definecolor{webgreen}{rgb}{0,.5,0}
\definecolor{webbrown}{rgb}{.6,0,0}
\definecolor{grigio}{rgb}{.85,.85,.85} 
\definecolor{RoyalBlue}{rgb}{0.0, 0.14, 0.4}
\definecolor{skyblue1}{rgb}{0.45,0.62,0.81}
\definecolor{skyblue2}{rgb}{0.2,0.39,0.64}
\definecolor{skyblue3}{rgb}{0.13,0.29,0.53}
\definecolor{scarlet1}{rgb}{0.93,0.16,0.16}
\definecolor{scarlet2}{rgb}{0.8,0,0}
\definecolor{scarlet3}{rgb}{0.64,0,0}

\definecolor{g}{gray}{0.50}

\hypersetup{
    colorlinks=true, linktocpage=true, pdfstartpage=1, pdfstartview=FitV,
    breaklinks=true, pdfpagemode=UseNone, pageanchor=true, pdfpagemode=UseOutlines,%
    plainpages=false, bookmarksnumbered, bookmarksopen=true, bookmarksopenlevel=1,%
    hypertexnames=true, pdfhighlight=/O,
    urlcolor=webbrown, linkcolor=RoyalBlue, citecolor=webgreen, 
    pdftitle={},%
    pdfauthor={Timur Aslyamov},%
    pdfsubject={},%
    pdfkeywords={},%
    pdfcreator={pdfLaTeX},%
    pdfproducer={LaTeX REVTeX}%
}

\begin{document}


\title{Faradaic and capacitive charging of an electrolyte-filled pore in response to a small applied potential}

\author{Timur Aslyamov}
\email{timur.aslyamov@uni.lu}
\affiliation{Department of Physics and Materials
Science, University of Luxembourg, L-1511 Luxembourg City, Luxembourg}

\author{Massimiliano Esposito}
\email{massimiliano.esposito@uni.lu}
\affiliation{Department of Physics and Materials
Science, University of Luxembourg, L-1511 Luxembourg City, Luxembourg}

\author{Mathijs Janssen}
\email{mathijs.a.janssen@nmbu.no}
\affiliation{Norwegian University of Life Sciences, Faculty of Science and Technology, Ås, Norway}

\date{\today}

\begin{abstract}
Electrochemical devices often charge both through Faradaic reactions and electric double layer formation.
Here, we study these coupled processes in a model system of a long electrolyte-filled pore subject to a small suddenly-applied potential, close to the equilibrium potential $\Psi^\text{eq}$ at which there is no net Faradaic charge transfer.
Specifically, we solve the coupled Poisson-Nernst-Planck and Frumkin-Butler-Volmer equations by asymptotic approximations, using the pore's small inverse aspect ratio as the small parameter.
In the early-time limit, the reaction–diffusion equations yield an extended Faradaic transmission line model that includes a voltage source, $\Psi_\text{eq}$, biasing the Faradaic reactions, captured by the resistance $R_F$. 
In the long-time limit, the model exhibits a nontrivial potential of zero charge, $\Psi_\text{pzc} = \Psi_\text{eq}[1 - \hat{Z}(0)/R_F]$, where $\hat{Z}(0)$ is the experimentally accessible zero-frequency impedance of the system. 
This expression provides a new means to experimentally measure the Faradaic contribution to $\Psi_\text{pzc}$.
\end{abstract}

\maketitle

\section{Introduction}
Electrochemistry deals with charge-transfer reactions across electrode-electrolyte interfaces.
At such interfaces, Faradaic charge transfer often goes along with non-Faradaic screening of electronic charge on the electrode through ionic charge in the electrolyte, known as the \textit{electric double layer (EDL)}. 
Concurrent Faradaic and non-Faradaic charging occurs in pseudocapacitors \cite{conway1991transition,simon2020perspectives,choi2020achieving}, corrosion \cite{zhang2024molecular,leger1999front}, electrochemical catalysis \cite{zhang2018progress,ringe2020double}, water treatment~\cite{alkhadra2022electrochemical,he2018theory,he2021theory}, and electrodes with defects or surface modifications \cite{evlashin2019n,evlashin2020role,bondareva2024insight}. 
As many of these examples involve porous electrodes, understanding concurrent Faradaic and non-Faradaic charging in confinement is fundamentally important to electrochemistry.
Historically, the two main tools to get such understanding have been effective circuits like the \textit{transmission line (TL)} model and the macrohomogenous approach of Newman and coworkers \cite{newman1962theoretical,newman1975porous,doyle1993modeling}.

\begin{figure*}
    \centering
    \includegraphics[width=\linewidth]{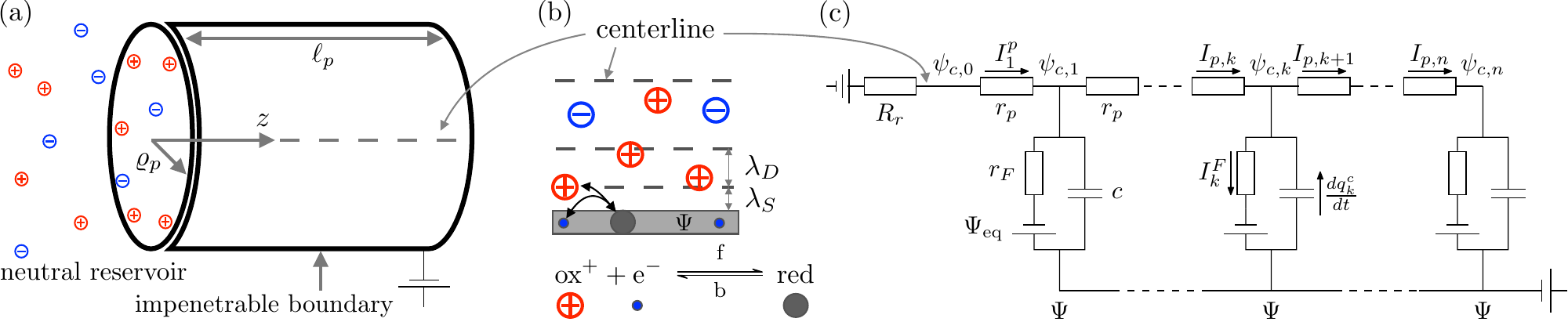}
    \caption{(a) Schematic of a cylindrical pore connected with a bulk reservoir. 
    The pore surface is chemically active with oxidation reaction \cref{eq:redox}.
    (b) Schematic illustration of the ions (red and blue discs are cations and anions, respectively) in a two-dimensional cut of the cylindrical setup colored gray in (a).  
    The pore obtains positive charge while the reservoir remains neutral. 
    (c) Faradaic TL circuit.}
    \label{fig:fig-1}
\end{figure*}

First, the TL model \cite{ksenzhek1956,danielbek1948,posey1966theory,keiser1976abschatzung,janssen2021transmission} captures ion transport and EDL formation in a long pore through a network wherein the total pore resistance $R_p$ and capacitance $C$ are distributed over infinitesimal resistors and capacitors.
De Levie included charge transfer through resistors with Faradaic resistance $R_F$ parallel to the capacitors \cite{levie1967electrochemical}; we refer to this extended circuit as the Faradaic TL model.
\Cref{fig:fig-1}(c) shows such a Faradaic TL circuit, which, unlike the ones in  \cite{levie1967electrochemical,pedersen2023equivalent,bumberger2024transmission}, contains a source of voltage $\Psi^\text{eq}$ biasing Faradaic reactions.
Away from the equilibrium potential, heterogeneous reactions create nontrivial potential and concentration profiles along the pore's centerline.
In turn, these profiles affect the overpotential and charge transfer resistances, which then vary along the pore even at steady state.
Lasia included these effects in several extended Faradaic TL models \cite{lasia2014electrochemical}.

Second, the macrohomogenous approach treats porous electrodes like a continuum, often of lower dimensionality, where pore and electrode phases coexist at each point, and where transport equations with effective parameters govern ionic and electronic transport.
Paasch and coworkers \cite{paasch1993theory} used this approach, with charge transfer included through the \textit{Butler-Volmer (BV)} equation, to determine a porous electrode's impedance.
A similar model by Devan and coworkers also included the spatiotemporal variation of ionic species, driven by diffusion and reactions \cite{devan2004analytical}.
Compared to \cite{paasch1993theory}, inclusion of the diffusive charge transport led, in the Nyquist representation, to an impedance curve with one more low-frequency arc.
Biesheuvel, Fu, and Bazant used the macrohomogenous approach to describe a porous electrode's transient response  \cite{biesheuvel2011diffuse}.
For the first time in this context, they used Frumkin's correction to the BV equation [giving the \textit{Frumkin-Butler-Volmer (FBV)} equation, \emph{viz}.~\cref{eq:FBV-def}], which says that charge transfer happens at the \textit{outer Helmholtz plane (OHP)}.
Accordingly, charge transfer is driven by the potential drop from the electrode to that plane, rather than by the potential drop between the electrode and a faraway point in the bulk electrolyte, as in the BV equation.
The Frumkin correction accounts better for the local reaction environment and leads to a consistent description of redox processes \cite{bazant2013theory,bazant2017thermodynamic}.

Despite the successes of equivalent circuit modeling and the macrohomogenous approach, both methods come with limitations.
Both method's coarse-grained starting points inherently lack information on the spatiotemporal charging of individual pores.
Details and effects can be added post-hoc, as was done for instance by Biesheuvel and coworkers who added Frumkin's correction to charge transfer \cite{biesheuvel2011diffuse}.
But less coarse, microscopic electrolyte models are more transparent in their assumptions and restrictions, and are therefore more straightforwardly extended.
In recent years, microscopic modeling has clarified TL model's region of validity and its underlying assumptions \cite{mirzadeh2014enhanced,bi2020molecular, henrique2021charging,henrique2022impact,yang2022simulating,pedersen2023equivalent}.
For example, two of us used the \textit{Poisson-Nernst-Planck (PNP)} equations to determine the evolution of the ion densities and electrostatic potential during the charging of a pore with a blocking surface \cite{aslyamov2022analytical}.
For the case of a long pore, thin EDLs, equal cationic and anionic diffusivities, and small applied potentials, we found an expression for the potential drop between the pore's surface and its centerline [Eq.~(44) there]---depending only on electrolyte properties (Debye length $\lambda_D$ and diffusivity $D$) and the pore's size and shape.
That equation was of identical form as predicted by the TL model, which, however, contained the lumped parameters $R_p$ and $C$.
By equating the PNP and TL results for the potential drop, we found expressions for $R_p$ and $C$ in terms of the microscopic parameters, which agreed with ad-hoc estimates thereof using a dilute electrolyte's resistivity and the Helmholtz EDL capacitance.
Hence, this analysis proved that, under the given restrictions, TL and PNP predictions coincided.
A major advantage of microscopic modeling is that it gives a transparent and straightforward---though often tedious---route to relax the restrictions.
For short pores \cite{pedersen2023equivalent}, overlapping EDLs \cite{henrique2022charging}, different diffusivities \cite{henrique2025parallel}, and larger potentials \cite{aslyamov2022analytical}, PNP models revealed that charging cannot be captured by standard TL models.
If at all, charging of these pores is only reproduced by circuits that are exotic to an extent it would be hard to come up with them (let alone justify) by eyeballing the underlying physics.   
 
Beside verifying previous course-grained models, and spelling out underlying assumptions, microscopic PNP modeling gives spatial information to pore charging not accessible from either macrohomogeneous approach or TL modeling. 
For instance, the PNP modeling inherently captures surface conduction, studied numerically by \cite{mirzadeh2014enhanced} and included ad-hoc in their TL model.
Our analytical PNP model \cite{aslyamov2022analytical} reproduced numerically-determined charging times from \cite{mirzadeh2014enhanced} without any fit parameters. 
The spatiotemporal information offered by microscopic models will become more relevant now that electrodes with well-ordered pore and channel structures can synthesized \cite{yoo2011ultrathin,yang2013liquid,yang2017mxene,he2020effects}, and spatiotemporal potentials can be mapped \cite{kutbay}.

In this article, we extend our PNP analysis of pore charging with charge transfer, modeled through the FBV equation. 
The PNP and FBV equations have already been solved analytically for electrolytes between oppositely charged flat electrodes \cite{bonnefont2001analysis,bazant2005current,chu2005electrochemical,van2010diffuse,vansoestbergen2010diffuse,yan2017theory,li2022impedance,Jarvey_2022,yan2021adaptive,barbero2022kinetic} and at charged solid-liquid interfaces and films \cite{levey2022finite}; numerical simulations of charging in pores were presented in \cite{butt2024local,he2021theory}. 
But such microscopic models for simultaneous Faradaic and non-Faradaic charging have not been analyzed analytically for porous electrodes charging.
Here, we analytically solve the FBV-PNP equations for a porous electrode model consisting of a single long pore, subject to a small applied potential and close to equilibrium. 
Importantly, we consider asymmetric kinetics of the chemical reaction, which leads to a nonzero equilibrium potential, $\Psi_\text{eq} \neq 0$, under the condition of zero electron flux. 
For early times, the linear response near $\Psi_\text{eq}$ turns out to be identical to that of the Faradaic TL circuit in \cref{fig:fig-1}(c), with distributed capacitance $C$, pore electrolyte resistance $R_p$, and Faradaic resistance $R_F$. 
As in \cite{aslyamov2022analytical}, where we studied blocking electrode charging, we now find expressions for the lumped circuit parameters in terms of microscopic pore properties, again, in complete agreement with earlier expressions \cite{lasia1995impedance}, providing a first-principles check of these expressions. 
Finally, we study the influence of redox reactions on the pore's \textit{potential of zero charge (PZC)}, and obtain the remarkably simple expression $\Psi_\text{pzc} = \Psi_\text{eq}[1 - \hat{Z}(0)/R_F]$, where $\hat{Z}(0)$ is the zero-frequency impedance, which could be found from extrapolation of experimental impedance measurements. 
So far, PZC is mostly measured experimentally as the minimum of differential capacitance \cite{lockett2008differential,ojha2020double} or using optical methods.
Our expressions allows one to determine the PZC in a new way. 

The paper is organized as follows. 
\Cref{sec:FBVsetup} presents the setup and governing equations.
\Cref{sec:assymptoticanalysis} presents an asymptotic analysis of the model to leading order in the pore's aspect ratio.
This yields a closed set of equations for the chemical potentials in the pore.
In \cref{sec:near_eq_chaging}, we restrict our attention to charging close to equilibrium.
\Cref{sec:faradiacTLmodel} shows how the early-time behavior of the pore is characterized by a TL circuit.
\Cref{sec:steady_state} discusses the system's steady state.
\Cref{sec:PZC} discusses our model's impedance and its nontrivial PZC.
The Discussion section \ref{sec:discussion} compares our results to prior work on porous electrode charging.
We conclude the paper in \cref{sec:conclusions}.

\section{FBV-PNP model for pore charging}\label{sec:FBVsetup}

\subsection{Setup}\label{sec:setup}
Consider a cylindrical pore of length $\ell_p$ and radius $\varrho_p$ with conducting and impenetrable walls, see \cref{fig:fig-1}(a).
We use a cylindrical coordinate system $(r,\theta,z)$ with $r$ the radial distance, $\theta$ the azimuthal angle, and $z$ the axial coordinate.
The pore is closed at $z= \ell_p$ and open at $z=0$, where it is in contact with a reservoir filled with a $1:1$ electrolyte of salt concentration $c_0$.
The pore is subject to a potential $\Psi$ with respect to a plane far away in the reservoir where the potential is zero.

We consider a case where the cations are the only electroactive species and, following Frumkin, we assume Faradaic reactions to happen at the OHP, at $r=\varrho_S\equiv\varrho_p - \lambda_S$, a Stern (S) layer's distance  from the pore's centerline, see \cref{fig:fig-1}(b).
The Stern layer of width $\lambda_S$ is a charge-free region next to the pore's surface, which accounts for the fact that the charge of (hydrated) ions cannot approach the pore's surface arbitrarily closely.
Otherwise, we ignore the solvent's and ion's finite sizes in the PNP equations \eqref{eq:general_model} below, which means that they are much smaller than the smallest geometric length scale, that is, the pore's radius. 
Below, we will study a long pore ($\ell_p\gg \varrho_p$) with a thin EDL ($\varrho_p\gg \lambda_D$) and Stern layer ($\varrho_p\gg \lambda_S$).
Initially, the pore will be at rest and subject to the equilibrium (eq) potential $\Psi^\text{eq}$, that is, the applied potential $\Psi$ for which there is no net electric current from Faradaic reactions, or equivalently, the open circuit potential.
Then, at $t=0$, we will step $\Psi$ slightly away from $\Psi^\text{eq}$, and study the resulting transient ionic and electric response. 
We treat the  electrode's surface as a perfect conductor and ignore electroconvection, which is reasonable for small applied potentials \cite{ratschow2024convection}.

\subsection{Governing equations}
Consider a one-step, one-electron oxidation-reduction reaction,
\begin{equation}
    \label{eq:redox}
    \text{ox}^{+} +\text{e}^-\ch{ <=>[ $f$ ][ $b$ ]}\text{red}\,,
\end{equation}
in forward ($f$) and backward ($b$) directions.
The oxidation reactant $\text{ox}^{+}$ is the electrolyte's cation, $\text{e}^-$ denotes the electrons, and the reduction product corresponds to the solid-electrode atoms, see \cref{fig:fig-1}(b). 
We model the reaction flux  
$\mathcal{J}_\text{rct} = \mathcal{J}_\text{f} - \mathcal{J}_\text{b}$ (unit \si{\per\second}), containing forward and backward fluxes, through the FBV equation 
\cite{van2010diffuse,vansoestbergen2010diffuse,bazant2013theory,bazant2017thermodynamic}, 
\begin{subequations}\label{eq:FBV-def}
  \begin{align}
    \mathcal{J}_{f}(t,z) &=k_f \rho_+(t, \varrho_{S},z)  \exp\left\{-\alpha [\Phi-\phi(t,\varrho_S, z)]\right\}\,,\\
    \mathcal{J}_{b}(t,z)&=
    k_b    \exp\left\{(1-\alpha)[\Phi-\phi(t,\varrho_S, z)]\right\}\,,
\end{align}  
\end{subequations}
where $k_f$ and $k_b$ are forward and backward rate constants (unit \si{\per\second}) and where, following \cite{van2010diffuse,biesheuvel2011diffuse}, we set the transfer coefficients to $\alpha=1/2$.  
Moreover, $\phi(t,r,\theta,z) = \psi e/(k_B T)$ is the dimensionless potential, with $\psi$ the electrostatic potential, $k_B T$ the thermal energy, and $e$ the proton charge.
Likewise, $\Phi=e\Psi/k_b T$ is the dimensionless applied potential, such that $\Phi(t)=\phi(t,\varrho_p,z)$.
As the Faradaic redox reaction \eqref{eq:redox} converts cations into electrode atoms, our results apply to cathodes, where cations react at the electrode surface. 
Therefore, in what follows, we consider $\Phi \leq 0$. 
In our model, we neglect the spatial growth of the electrode when cations are deposited; the pore geometry is not affected by \cref{eq:redox}.
In \cref{eq:FBV-def}, $\phi(t,r,\theta,z)$ and the dimensionless ion densities $\rho_{\pm}= c_{\pm}/c_0$ are evaluated at $\varrho_s$, hence, \cref{eq:redox} asserts that Faradaic reactions happen at the OHP, see \cref{fig:fig-1}b.  

We model $\rho_\pm$ and $\phi$ in the electrolyte by the PNP equations,
\begin{subequations}\label{eq:general_model}
\begin{align}
    \label{eq:general_model-plus}
    \partial_t \rho_+&=\bm{\nabla}\cdot \boldsymbol{j}_{+} - \mathcal{J}_\text{rct}\delta\left(\frac{r}{\varrho_{S}}-1\right)\,,\\
    \label{eq:general_model-minus}
     \partial_t \rho_-&=\bm{\nabla}\cdot \boldsymbol{j}_{-}\,,\\
    \label{eq:general_model-flux}
    \boldsymbol{j}_{\pm}&=D\rho_{\pm}\bm{\nabla}\mu_{\pm}\,, \\
    \label{eq:general_model_mu}
    \mu_\pm&=\log(\rho_\pm )\pm \phi\,, \\
    \label{eq:general_model_Poisson}
    -\nabla^2\phi&=\frac{\rho_+-\rho_-}{2\lambda_D^2}\,,
\end{align}
\end{subequations}
where $D$ is the diffusion coefficient (unit \si{\meter\squared\per\second}), assumed spatially constant and the same for both ion species, $\mu_{\pm}$ are the dimensionless ionic chemical potentials (the chemical potentials scaled to $k_B T$), and $\lambda_D=[\varepsilon k_B T/(2c_0 e^2)]^{1/2}$ is the Debye length, with $\varepsilon$ the permittivity.
Within the Stern layer, $\varrho_S<r<\varrho_p$, all ionic densities are zero, so \cref{eq:general_model_Poisson} reduces to $\nabla^2\phi = 0$ there.
Moreover, as the pore's initial and boundary conditions [cf. \cref{sec:initialandboundaryconditions}] are rotationally symmetric, the ionic concentrations $\rho_{\pm}(t,r,z)$ and electrostatic potential $\psi(t,r,z)$ do not depend on $\theta$.
Accordingly, in \cref{eq:general_model} $\bm{\nabla}=(\partial_r,\partial_z)^\intercal$ is the 2d gradient and $\boldsymbol{j}_\pm = (j_{\pm,r}, j_{\pm,z})^\intercal$ ionic fluxes (unit \si{\meter\per\second}).

Last, the Dirac delta function term in \cref{eq:general_model-plus} accounts the heterogeneous redox reactions including Frumkin's correction and acts as a source term for the cationic number density $\rho_+(t,r,z)$.
Other works \cite{bonnefont2001analysis,van2010diffuse,vansoestbergen2010diffuse,li2022impedance,Jarvey_2022,he2021theory} implemented such heterogeneous reactions through boundary conditions on the current; we implement them here into the governing equations. 

\subsection{Initial and boundary conditions}\label{sec:initialandboundaryconditions}
Up to \cref{sec:linearchargingdynamics}, the only constraint we will put on the initial ionic densities $\rho_\pm^\text{ic}(\varrho,z)\equiv\rho_\pm(t=0,\varrho,z)$ is that they have rotational symmetry; otherwise, the PNP \cref{eq:general_model} would contain nontrivial fluxes in the $\theta$ direction as well.
From \cref{sec:linearchargingdynamics} onward, we will study a case where pore charging in response to an applied potential starts from equilibrium, that is, the state with no net electric current through the pore's surface [$\mathcal{J}_\text{rct}(z,t)=0$], and corresponding equilibrium potential $\Phi^\text{eq}$ and nonhomogeneous ion densities $\rho_\pm^\text{eq}(r)$.

The dimensionless potential is subject to the following boundary conditions,
\begin{subequations}\label{eq:bc-phi}
\begin{align}
    \phi(t,\varrho_p,z)&=\Phi\,,\label{eq:bc-phi_rp}\\
    \quad \partial_r\phi(t,0,z) &= 0\,,\label{eq:bc-phi_r0}
\end{align}
\end{subequations}
where the above-mentioned relation \eqref{eq:bc-phi_rp} can be seen as the definition of $\Phi$, and where \cref{eq:bc-phi_r0} follows from the rotational symmetry of $\phi(t, r, z)$.

For the ionic fluxes we have
\begin{subequations}
\label{eq:bc-j}
    \begin{align}
    \label{eq:bc-j-entrance}
        j_{\pm,z}(t,r,0)&\equiv \mathcal{L}\mu_{\pm}(t,r,0)\,,\\
        j_{\pm,r}(t,r,0)&\equiv 0\,,\\
    \label{eq:bc-j-end}
        j_{\pm,z}(t,r,\ell_p)&=0\,,\\
    \label{eq:bc-j_r}
        j_{\pm,r}(t,\varrho_p,z)&=0\,,\\
    \label{eq:bc-j_r2}
        j_{\pm,r}(t,0,z)&=0\,,    
    \end{align}
\end{subequations}
where \cref{eq:bc-j-end,eq:bc-j_r} are the boundary conditions for the $z$ and $r$ components of $\boldsymbol{j}$ on the impenetrable walls and where \cref{eq:bc-j_r2} follows from the rotational symmetry of the setup; For $z=0$, in \cref{eq:bc-j-entrance} we assume that the fluxes at the pore entrance can be modeled in terms of the Onsager theory with  $\mathcal{L}$ being the linear response coefficients. 
Notice that we only explicitly model the electrolyte dynamics in the pore, not in the bulk electrolyte reservoir.
In \cref{eq:bc-linear-regime-1}, we will establish that $\mathcal{L}\propto R_r^{-1}$, with $R_r$ representing the resistance of the reservoir; the charge influx is governed by Ohm's law.
In general, $R_r$ will depend on the geometry of the reservoir and the counter electrode, which we will not explicitly model here \cite{newman1966resistance, yang2022simulating,pedersen2023equivalent}. 

\section{Charging dynamics for a long pore subject to a small potential}\label{sec:assymptoticanalysis}
We use the pore's inverse aspect ratio $h=\varrho_p/\ell_p\ll1$ as the small parameter in an asymptotic analysis of the FBV-PNP equations \eqref{eq:FBV-def} and \eqref{eq:general_model} and their boundary conditions \eqref{eq:bc-phi} and \eqref{eq:bc-j}.
In doing so, we extend prior work that used asymptotic approximations to study EDL formation in blocking pores through the PNP equations \cite{aslyamov2022analytical,alizadeh2017multiscale} and dynamical density functional theory \cite{aslyamov2022relation,tomlin2022impedance}.

\subsection{Asymptotic approximation}
We define time scales $\tau_r=\varrho_p^2/D$ and $\tau_z= \ell_p^2/D$ characterizing the dynamics along the $r$-axis and $z$-axis, respectively, and dimensionless $z$- and $r$-coordinates by $\tilde{z}=z/L$ and $\tilde{r}=r/\varrho_p$.
Next, $\tilde{\phi}=\phi/\Phi$ is the scaled dimensionless potential and $\tilde{\rho}_{\pm} = \varrho_p^2/(2\Phi \lambda_D^2)\rho_\pm$ are the scaled concentrations, together yielding a scaled Poisson equation below [\cref{eq:nondim-Poisson}].
Finally, $\tilde{\mathcal{J}}_\text{rct}=\mathcal{J}_\text{rct}\varrho_p^2\tau_z/(2\Phi \lambda_D^2)$ is the dimensionless reaction flux.
We assume that the time scale of reaction dynamics is comparable to the time scale of the diffusion process along the pore. 
Therefore, the EDL in our theory will be in local equilibrium, unlike the nonequilibrium EDLs in Refs.~\cite{he2006dynamic,bier2024non}, who studied one-dimensional setups without time-scale separation.

These scaled variables enable us to identify how the various terms of \cref{eq:general_model} scale with $h$,
\begin{subequations}\label{eq:nondim_equations}
\begin{align}
    \label{eq:nondim-RD}
    &h^2\partial_{\tilde{t}} \tilde{\rho}_{+}=h^2\partial_{\tilde{z}}(\tilde{\rho}_+\partial_{\tilde{z}}\mu_+)+
    \frac{1}{\tilde{r}}\partial_{\tilde{r}}(\tilde{r}\tilde{\rho}_+\partial_{\tilde{r}}\mu_+)
    -h^2\tilde{\mathcal{J}}_\text{rct}\delta\left(\frac{r}{\varrho_{S}}-1\right)\,,\\
    \label{eq:nondim-RD-2}
    &h^2\partial_{\tilde{t}} \tilde{\rho}_{-}=h^2\partial_{\tilde{z}}(\tilde{\rho}_-\partial_{\tilde{z}}\mu_-)+
    \frac{1}{\tilde{r}}\partial_{\tilde{r}}(\tilde{r}\tilde{\rho}_-\partial_{\tilde{r}}\mu_-)\,,\\
    \label{eq:nondim-Poisson}
    &-\frac{1}{\tilde{r}} \partial_{\tilde{r}}(\tilde{r}\partial_{\tilde{r}}\tilde{\phi})-h^2\partial_{\tilde{z}}^2\tilde{\phi}=\tilde{\rho}_+-\tilde{\rho}_-\,.
\end{align}
\end{subequations}
We insert the following asymptotic expansions \footnote{As \cref{eq:nondim_equations} only contains even powers of $h$, we insert asymptotic expansions with even powers only as well, as uneven powers do not give new information.}, 
\begin{subequations}\label{eq:asymptotic_expansion}
\begin{align}
    \tilde{\rho}_\pm &= \tilde{\rho}^0_\pm + h^2 \tilde{\rho}^1_\pm + \mathcal{O}\left(h^4\right)\,, \\
    \tilde{\phi} &= \tilde{\phi}^0 + h^2 \tilde{\phi}^1 + \mathcal{O}\left(h^4\right)\,,\\
    \mu_{\pm} &= \mu_{\pm}^0 + h^2 \mu_{\pm}^1 + \mathcal{O}\left(h^4\right)\,,\\
    \tilde{\mathcal{J}}_\text{rct} &= \tilde{\mathcal{J}}_\text{rct}^0 + h^2 \tilde{\mathcal{J}}_\text{rct}^1 + \mathcal{O}\left(h^4\right)\,,
\end{align}
\end{subequations}
in \cref{eq:nondim_equations} and expand the dynamical equations \eqref{eq:nondim-RD} and \eqref{eq:nondim-RD-2} up to $\mathcal{O}(h^2)$ and the Poisson equation \eqref{eq:nondim-Poisson} up to $\mathcal{O}(h^0)$ 
\begin{subequations}\label{eq:nondim_equations-details}
\begin{align}
    \label{eq:nondim-RD-details}
    &h^2\partial_{\tilde{t}} \tilde{\rho}^0_{+}=h^2\partial_{\tilde{z}}(\tilde{\rho}^0_+\partial_{\tilde{z}}\mu_+^0)+
    \frac{1}{\tilde{r}}\partial_{\tilde{r}}(\tilde{r}\tilde{\rho}^0_+\partial_{\tilde{r}}\mu^0_+)\nn
    &\qquad\qquad
    +\frac{h^2}{\tilde{r}}\partial_{\tilde{r}}[\tilde{r}(\tilde{\rho}^1_+\partial_{\tilde{r}}\mu^0_++\tilde{\rho}^0_+\partial_{\tilde{r}}\mu^1_+)]-h^2\tilde{\mathcal{J}}^0_\text{rct}\delta\left(\frac{r}{\varrho_{S}}-1\right)
    \nn
    &\qquad\qquad+\mathcal{O}(h^4)\,,\\
    \label{eq:nondim-RD-2-details}
    &h^2\partial_{\tilde{t}} \tilde{\rho}^0_{-}=h^2\partial_{\tilde{z}}(\tilde{\rho}^0_-\partial_{\tilde{z}}\mu_-^0)+
    \frac{1}{\tilde{r}}\partial_{\tilde{r}}(\tilde{r}\tilde{\rho}^0_-\partial_{\tilde{r}}\mu^0_-)
    \nn
    &\qquad\qquad+\frac{h^2}{\tilde{r}}\partial_{\tilde{r}}[\tilde{r}(\tilde{\rho}^1_-\partial_{\tilde{r}}\mu^0_-+\tilde{\rho}^0_-\partial_{\tilde{r}}\mu^1_-)]+\mathcal{O}(h^4)\,,\\
    \label{eq:nondim-Poisson-details}
    &-\frac{1}{\tilde{r}} \partial_{\tilde{r}}(\tilde{r}\partial_{\tilde{r}}\tilde{\phi}^0)=\tilde{\rho}^0_+-\tilde{\rho}^0_-+\mathcal{O}(h^2)\,.
\end{align}
\end{subequations}
Likewise, we insert \cref{eq:asymptotic_expansion} into the boundary conditions \eqref{eq:bc-phi},
\begin{subequations}\label{eq:nondim_boundary_expansion}
\begin{align}
    \label{eq:nondim_boundary_expansion_Phi}
    \tilde{\phi}^0(\tilde{t},1,\tilde{z})&=1\,,\\
    \label{eq:nondim_boundary_expansion_phi_end}
    \partial_{\tilde{r}}\tilde{\phi}^0(\tilde{t},0,\tilde{z})&=0\,,
\end{align}
\end{subequations}
and \eqref{eq:bc-j},
\begin{subequations}\label{eq:nondim_boundary_expansion}
\begin{align}    
    j^0_{\pm,z}(t,r,0)=\mathcal{L}\mu^0_{\pm}&(t,r,0)\,,\\
    j^0_{\pm,z}(t,r,\ell_p)&=0\,,\\
    \label{eq:nondim_boundary_expansion_jr}
    \tilde{\rho}^0\partial_{\tilde{r}}\mu_\pm^0\Big|_{\tilde{r}=1} + h^2\left(\tilde{\rho}^1_\pm\partial_{\tilde{r}}\mu_\pm^0 +\tilde{\rho}^0_\pm\partial_{\tilde{r}}\mu_\pm^1\right)\Big|_{\tilde{r}=1} &= 0\,,\\
        \label{eq:nondim_boundary_expansion_jr2}
    \tilde{\rho}^0\partial_{\tilde{r}}\mu_\pm^0\Big|_{\tilde{r}=0} + h^2\left(\tilde{\rho}^1_\pm\partial_{\tilde{r}}\mu_\pm^0 +\tilde{\rho}^0_\pm\partial_{\tilde{r}}\mu_\pm^1\right)\Big|_{\tilde{r}=0} &= 0\,,
\end{align}
\end{subequations}
where $j^0_{\pm,z}=\rho^0_\pm\partial_z \mu^0_\pm$.
As in \cite{aslyamov2022analytical,aslyamov2022relation}, by collecting terms with the same order of $h$ in \cref{eq:nondim_equations-details} we find equations for $\tilde{\rho}^0_\pm(t,z)$ and $\tilde{\phi}^0(t,z)$. 
Following these papers, we do not aim to solve our system of equation up to $O(h^2)$ but, expanding to $O(h^2)$, we find a closed system of equations [\eqref{eq:asymptotic-RD}, \eqref{eq:asymptotic-RD-conditions}, and \eqref{eq:rho-aver-explicit}] for the cross sectional averaged first order variables ($\overline{\rho}^0_\pm,\overline{\phi}^0$, and $\overline{\mu}^0_\pm $).
Here, bars denote the cross-sectional averages of a variable $f(t,r,\theta,z)$, 
\begin{equation}\label{eq:cross-sectional-average}
    \overline{f}(t,z)=\frac{1}{A_p}\int_{0}^{\varrho_p}\int_{0}^{2\pi} f(t,r,\theta, z)\, r\,dr\, d\theta \,,
\end{equation}
with $A_p=\pi \varrho_p^2$ the pore's cross sectional area.

\subsection{Local cross-sectional equilibrium}\label{sec:poisson}
At $\mathcal{O}(h^0)$, \cref{eq:nondim-RD-details} reads $(1/\tilde{r})\partial_{\tilde{r}}(\tilde{r}\tilde{\rho}_\pm^0\partial_{\tilde{r}}\mu_\pm^0)=0$. 
With the boundary conditions \eqref{eq:nondim_boundary_expansion_jr} and \eqref{eq:nondim_boundary_expansion_jr2} at that order, $\partial_{\tilde{r}}\mu_{\pm}^0(\tilde{t},1,\tilde{z})=0$ and $\partial_{\tilde{r}}\mu_{\pm}^0(\tilde{t},0,\tilde{z})=0$, we find
\begin{equation}
    \label{eq:asymptotic-equilibrium}
    \partial_r\mu^0_\pm=0 \rightarrow \mu_\pm(t,r,z) = \mu^0_\pm(t, z)+\mathcal{O}(h^2)\,,
\end{equation}
showing that, up to $\mathcal{O}(h^2)$, the chemical potential is constant in the radial direction for all $z$.
Hence, while $\mu_\pm(t, z)$ varies with $t$ and $z$, each cross section is in local equilibrium. 

\subsection{Cross-section averaged dynamics}\label{sec:dynamics}
Next, at $\mathcal{O}(h^2)$, \cref{eq:nondim-RD-details,eq:nondim-RD-2-details} read
\begin{subequations}\label{eq:nondim_equations-h2}
\begin{align}
\label{eq:nondim-RD-h2}
    \partial_{\tilde{t}} \tilde{\rho}^0_{+}&=\partial_{\tilde{z}}(\tilde{\rho}^0_+\partial_{\tilde{z}}\mu^0_+)+\frac{1}{\tilde{r}}\partial_{\tilde{r}}(\tilde{r}\tilde{\rho}_+^{1}\partial_{\tilde{r}}\mu^0_+)+\frac{1}{\tilde{r}}\partial_{\tilde{r}}(\tilde{r}\tilde{\rho}^0_+\partial_{\tilde{r}}\mu_+^{1})\nn
    &\quad - \tilde{\mathcal{J}}^0_\text{rct}\delta\left(\frac{r}{\varrho_{S}}-1\right)\,,\\
\label{eq:nondim-RD-2-h2}
    \partial_{\tilde{t}} \tilde{\rho}^0_{-}&=\partial_{\tilde{z}}(\tilde{\rho}^0_-\partial_{\tilde{z}}\mu^0_-)+
    \frac{1}{\tilde{r}}\partial_{\tilde{r}}(\tilde{r}\tilde{\rho}_-^{1}\partial_{\tilde{r}}\mu^0_-)+
    \frac{1}{\tilde{r}}\partial_{\tilde{r}}(\tilde{r}\tilde{\rho}^0_-\partial_{\tilde{r}}\mu_-^{1})\,.
\end{align}
\end{subequations}
Taking cross-sectional averages [\cref{eq:cross-sectional-average}], we find
\begin{subequations}\label{eq:nondim_equations-h2-2}
\begin{align}
\label{eq:nondim-RD-h2-2}
    \partial_{\tilde{t}} \overline{\tilde{\rho}}^0_{+}&=\partial_{\tilde{z}}(\overline{\tilde{\rho}}^0_+\partial_{\tilde{z}}\mu^0_+) - 2\tilde{\mathcal{J}}^0_\text{rct}
    +2\underbrace{\tilde{r}\left(\tilde{\rho}_+^{1}\partial_{\tilde{r}}\mu^0_+
    +\tilde{\rho}^0_+\partial_{\tilde{r}}\mu_+^{1}\right)\Big|_{\tilde{r}=0}^{\tilde{r}=1}}_{=0}\,,\\
\label{eq:nondim-RD-2-h2-2}
    \partial_{\tilde{t}} \overline{\tilde{\rho}}^0_{-}&=\partial_{\tilde{z}}(\overline{\tilde{\rho}}^0_-\partial_{\tilde{z}}\mu^0_-)
    +2\underbrace{\tilde{r}\left(\tilde{\rho}_-^{1}\partial_{\tilde{r}}\mu^0_-
    +\tilde{\rho}^0_-\partial_{\tilde{r}}\mu_-^{1}\right)\Big|_{\tilde{r}=0}^{\tilde{r}=1}}_{=0}\,,
\end{align}
\end{subequations}
where for the second term in \cref{eq:nondim-RD-h2-2} we used
\begin{align}
    &\frac{1}{A_p}\int_{0}^{\varrho_p}\int_{0}^{2\pi} \delta\Big(\frac{r}{\varrho_S}-1\Big)\, r\,dr\, d\theta \\ \nonumber
    &= \frac{2\pi\varrho_S^2}{A_p}\underbrace{\int_{0}^{\varrho_{p}/\varrho_{S}}\delta(u-1)u\,du}_{=1}\approx 2\,,
\end{align}
with approximation sign due to $\varrho_p\approx\varrho_S$.

In \cref{eq:nondim_equations-h2-2}, the terms with braces drop because of the boundary conditions \eqref{eq:nondim_boundary_expansion_jr} and \eqref{eq:nondim_boundary_expansion_jr2} on the radial flux, $\tilde{\rho}^0\cancel{\partial_{\tilde{r}}\mu_\pm^0}\big|_{\tilde{r}=1} + h^2\left(\tilde{\rho}^1_\pm\partial_{\tilde{r}}\mu_\pm^0 +\tilde{\rho}^0_\pm\partial_{\tilde{r}}\mu_\pm^1\right)\big|_{\tilde{r}=1} = 0$, where the first term drops because $\mu^0_{\pm}(t,z)$ does not depend on $r$ [\cref{eq:asymptotic-equilibrium}].

Returning to the unscaled variables, we rewrite  \cref{eq:nondim_equations-h2-2} to 
\begin{subequations}\label{eq:asymptotic-RD}
\begin{align}
\label{eq:asymptotic-RD-plus}
    \partial_{t} \overline{\rho}^0_{+}&=D\partial_{z}(\overline{\rho}^0_{+}\partial_{z}\mu^0_+)-2\mathcal{J}_\text{rct}(\mu^0_\pm, \Phi)\,,\\
\label{eq:asymptotic-RD-minus}
    \partial_{t} \overline{\rho}^0_{-}&=D\partial_{z}(\overline{\rho}^0_{-}\partial_{z}\mu^0_-)\,,
\end{align}
\end{subequations}
which are subject to boundary and initial conditions that follow from cross-sectional averages of \cref{eq:nondim_boundary_expansion}, 
\begin{subequations}\label{eq:asymptotic-RD-conditions}
\begin{align}
\label{eq:asymptotic-RD-bc-1}
    D\overline{\rho}^0_{\pm}(t,0)\partial_z\mu^0_\pm(t,0)&=\mathcal{L}\mu^0_{\pm}(t,0)\,,\\
\label{eq:asymptotic-RD-bc-2}
    \partial_z\mu^0_\pm(t,\ell_p)&=0\,,\\
    \overline{\rho}^0_{\pm}(0,z)&=\overline{\rho}_{\pm}^{0,\text{ic}}(z)\,.\label{eq:asymptotic-RD-bc-3}
\end{align}
\end{subequations}   
\Cref{eq:asymptotic-RD} coincides with the reaction-diffusion equations of nonideal mixtures \cite{aslyamov2023nonidel,avanzini2024nonequilibrium}.
Beside the last term in \cref{eq:asymptotic-RD-plus}, \cref{eq:asymptotic-RD} is identical to Eq.~(8) in our work on the charging of blocking pores \cite{aslyamov2022analytical}.
As in that article, we see that the lowest-order dynamics of the pore relaxation appears at $\mathcal{O}(h^2)$.
Moreover, after taking cross-sectional averages, the dynamical equations \eqref{eq:asymptotic-RD} and \eqref{eq:asymptotic-RD-conditions} only contain $\mathcal{O}(h^0)$ terms of the expansions \eqref{eq:asymptotic_expansion} ($\overline{\rho}^0_\pm,\overline{\phi}^0$, and $\overline{\mu}^0_\pm $).
Hence, to describe the pore's relaxation to lowest order in $h$, we can ignore $\overline{\rho}^1_\pm,\overline{\phi}^1$, and $\overline{\mu}^1_\pm$.
However, \cref{eq:asymptotic-RD,eq:asymptotic-RD-conditions} are not a set of closed equation yet, as we have not expressed $\mu$ in terms of $\overline{\rho}$. 
Using \cref{eq:general_model_mu}, we find
\begin{align}
\label{eq:closer-1}
    \overline{\rho}(t,z) = \frac{e^{\mu_\pm(t,z)}}{\varrho_p^2}\int e^{\mp\phi(t,r, z)} r \,d r\,,
\end{align}
and in the next subsection, we will derive $\overline{\rho}_\pm(t,z) = \overline{\rho}_\pm\bm{(}\mu_{\pm}(t,z)\bm{)}$ [\cref{eq:rho-aver-explicit}] to close \cref{eq:asymptotic-RD,eq:asymptotic-RD-conditions}.

\subsection{$\mathcal{O}(h^0)$ solution of the Poisson equation \eqref{eq:nondim-Poisson}}\label{sec:poisson}
At $\mathcal{O}(1)$, the Poisson equation \eqref{eq:nondim-Poisson-details} and its boundary conditions \cref{eq:nondim_boundary_expansion_Phi,eq:nondim_boundary_expansion_phi_end} read
\begin{subequations}\label{eq:asymptotic-Poisson}
\begin{align}
   -\frac{1}{\tilde{r}} \partial_{\tilde{r}}(\tilde{r}\partial_{\tilde{r}}\tilde{\phi}^0)&=\tilde{\rho}_+^0-\tilde{\rho}_-^0\,,\label{eq:asymptotic-Poisson-firstline}\\
    \tilde{\phi}^0(\tilde{t},1,\tilde{z})=1&\,,\quad\partial_{\tilde{r}}\tilde{\phi}^0(\tilde{t},0,\tilde{z})=0\label{eq:asymptotic-Poisson-bc}\,.
\end{align}
\end{subequations}

For brevity, from hereon we write $\tilde{\rho}_\pm = \tilde{\rho}^0_\pm(t,z)$, $\tilde{\phi} = \tilde{\phi}^0(t,z)$, and $\mu_\pm = \log\rho_{\pm}^0(t,z) \pm \phi^0(t,z)$ and return to non-scaled variables. 
Likewise, we will write  $\rho_\pm = \rho^0_\pm$, $\phi = \phi^0$, $\mu_\pm = \mu_\pm^0$, and $\mathcal{J}_\text{rct} = \mathcal{J}_\text{rct}^0$.

To determine the right-hand side of \cref{eq:asymptotic-Poisson-firstline}, we exploit the pore's cross-sectional equilibrium \eqref{eq:asymptotic-equilibrium} and rewrite \cref{eq:general_model_Poisson} to $\rho_{\pm}(t,r,z) = \exp[\mu_{\pm}(t,z)\mp\phi(t,r,z)]$.
For $|\Phi|\ll 1$, we can omit terms of order $\mathcal{O}(\phi^2)$, so 
\begin{equation}\label{eq:rho-linear}
    \rho_{\pm}(t,r,z) = e^{\mu_\pm(t,z)}\big[1\mp\phi(t,r,z)\big]\,,
\end{equation}
and
\begin{equation}\label{eq:charge-linear}
  \rho_{+}-\rho_{-}= \left(e^{\mu_+}+e^{\mu_-}\right)\left(\tanh m_--\phi\right)\,,
\end{equation}
with 
\begin{align}
\label{eq:m-def}
    m_\pm \equiv \frac{(\mu_+\pm\mu_-)}{2}\,.
\end{align}
Inserting \cref{eq:charge-linear} in \cref{eq:asymptotic-Poisson}, we arrive at a Debye-H\"{u}ckel-like equation,
\begin{subequations}\label{eq:Poisson-linear}
\begin{align}
   \frac{1}{r} \partial_{r}(r\partial_{r}\phi_m)&=\frac{\phi_m}{\lambda_m^2}\,,\\
    \phi_m(t,\varrho_S,z)=\Phi_m&\,,\quad \partial_{r}\phi_m(t,0,z)=0\,,\label{eq:Poisson-linear-bc}
\end{align}
\end{subequations}
where we shifted the boundary condition \eqref{eq:asymptotic-Poisson-bc} to the OHP and introduced modified variables,
\begin{subequations}\label{eq:modified-parameters}
\begin{align}
    \label{eq:phi-m}
    \phi_m &\equiv \tanh m_- -\phi(t,r,z)\,,\\
    \label{eq:Phi_m}
    \Phi_m & \equiv\tanh m_- - \phi(t,\varrho_S,z)\,,\\
    \label{eq:lambda_m}
    \lambda_m &\equiv\lambda_D\sqrt{\frac{2}{e^{\mu_+}+e^{\mu_-}}}\,.
\end{align}
\end{subequations}
\Cref{eq:Poisson-linear} is solved by $\phi_m=\Phi_m {\rm I}_0(r/\lambda_D)/{\rm I}_0(\varrho_S/\lambda_m)$, where ${\rm I}_k$ is the $k$th-order modified Bessel function of the first kind. 
With \cref{eq:phi-m,eq:Phi_m}, we find
\begin{equation}\label{eq:phi-cyl}
    \phi(t,r,z) = \tanh m_-  - \big[\tanh m_- - \phi(t,\varrho_S,z) \big]\frac{{\rm I}_0(r/\lambda_m)}{{\rm I}_0(\varrho_S/\lambda_m)}\,,
\end{equation}
for $r\leq \varrho_S$. 
Note that $\phi(t,r,z)$ depends only implicitly on $t$ and $z$, through $\mu_{\pm}(t,z)$ in $m_\pm$ and $\lambda_m$. 

To determine the potential $\phi(t,\varrho_S,z)$ at the OHP, we note that, in the Stern layer ($\varrho_S\leq r \leq\varrho_p$), the Poisson equation reduces to the Laplace equation, $\partial_r(r\partial_r\phi) = 0$.
Integrating twice gives $\phi = k_1\ln(r/\varrho_p)+k_2$, where $k_1$ and $k_2$ are integration constants.
Enforcing \cref{eq:nondim_boundary_expansion_Phi} yields $k_2=\Phi$.
Next, cross-sectional averaging of \cref{eq:asymptotic-Poisson} and employing the divergence theorem yield
\begin{align}
    \label{eq:bc-Stern-0}
    \frac{2\pi}{\pi\varrho_S^2}\int_{0}^{\varrho_S} \frac{1}{r}\partial_r (r\partial_r\phi) \,rdr &= - \frac{1}{2\lambda_D^2}(\overline{\rho}_+-\overline{\rho}_-)\,,\nonumber\\
    2\varrho_S\partial_r\phi\big|_{r=\varrho_{S}} &= -\frac{\varrho_S^2}{2\lambda_D^2}(\overline{\rho}_+-\overline{\rho}_-)\,,
\end{align}
giving $k_1 = -(\varrho_S/2\lambda_D)^2(\overline{\rho}_+ - \overline{\rho}_-)$.
The potential at the OHP is thus $\phi(t,\varrho_S,z) = - \varrho_p^2/(4\lambda_D^2)\ln(1-\lambda_S/\varrho_p)(\overline{\rho}_+-\overline{\rho}_-)+\Phi$, which, for $\lambda_S\ll\varrho_p$ reduces to
\begin{equation}\label{eq:bc-Stern2}
    \phi(t,\varrho_S,z) = \frac{\varrho_p \lambda_S}{4\lambda_D^2}(\overline{\rho}_+-\overline{\rho}_-)+\Phi\,,
\end{equation}
such that \cref{eq:phi-cyl} reads
\begin{align}
    \label{eq:phi-cyl-2}
    \phi(t,r,z) &= \tanh m_- \\
    &- \left(\tanh m_- - \frac{\varrho_p \lambda_S(\overline{\rho}_+-\overline{\rho}_-)}{4\lambda_D^2} - \Phi\right) \frac{{\rm I}_0(r/\lambda_m)}{{\rm I}_0(\varrho_S/\lambda_m)}\nonumber\,.
\end{align}

Next, we find $\overline{\rho}_+-\overline{\rho}_-$. 
Calculating the cross-section average [\emph{viz}.~\cref{eq:cross-sectional-average}] of  \cref{eq:charge-linear} we find
\begin{align}
\label{eq:q-equation}
    \overline{\rho}_+-\overline{\rho}_- &= \frac{4\lambda_D^2}{\lambda_m\varrho_p}\left(\tanh m_- - \frac{\varrho_p \lambda_S(\overline{\rho}_+-\overline{\rho}_-)}{4\lambda_D^2} - \Phi\right)\,,
\end{align}
where we approximated $\varrho_p\approx\varrho_S$ and where we used $2\lambda_D^2/\lambda_m^2=e^{\mu_+}+e^{\mu_-}$ and 
$\int_0^{\varrho_p} r {\rm I}_0(r/\lambda_m)\,dr=\lambda_m\varrho_p {\rm I}_1(\varrho_p/\lambda_m)$ with ${\rm I}_1(\varrho_p/\lambda_m)/{\rm I}_0(\varrho_p/\lambda_m)\approx 1$. Extracting $\overline{\rho}_+-\overline{\rho}_-$ from \cref{eq:q-equation} we find 
\begin{align}
\label{eq:q-explicit}
    \overline{\rho}_+-\overline{\rho}_- &= 2\left(\tanh m_--\Phi\right)\Lambda_m\,,
\end{align}
where introduced the dimensionless parameter $\Lambda_m \equiv 2\lambda_m^2/[(\lambda_m+\lambda_S)\varrho_p]$.
Inserting \cref{eq:q-explicit} into \cref{eq:phi-cyl-2} we find 
\begin{align}
\label{eq:phi-explicit}
   \phi\bm{(}\mu_\pm(t,z);r\bm{)} &= \tanh m_- -\left(1-\frac{\lambda_m^2\lambda_S}{\lambda_D^2(\lambda_m+\lambda_S)}\right) \nn
   & \qquad\times\left(\tanh m_- -\Phi\right) \frac{{\rm I}_0(r/\lambda_m)}{{\rm I}_0(\varrho_S/\lambda_m)}\,,
\end{align}
in terms of the time-dependent chemical potentials $\mu_{\pm}(t,z)$ that enter $m_-$ and $\lambda_m$.
Hence, once we know $\mu_{\pm}(t,z)$, we can reconstruct the potential $\phi(t,r,z)$ by \cref{eq:phi-explicit}. 
For the centerline (c) electrostatic potential $\phi_{c}(t,z)\equiv\phi(t,r=0,z)$, \cref{eq:phi-explicit} simplifies significantly as
\begin{align}
\label{eq:phi-c-tanh_m}
   \phi_c(t,z) &= \tanh m_-\,,
\end{align}
where we use ${\rm I}_0(0)/{\rm I}_0(\varrho_S/\lambda_m) \ll 1$ for $\varrho_S \gg \lambda_m$.

Finally integrating \cref{eq:rho-linear} with \cref{eq:phi-explicit}, we find  
\begin{align}
\label{eq:rho-aver-explicit}
    \overline{\rho}_{\pm}&=e^{\mu_\pm}\bigg[1\mp\tanh m_-\pm\frac{2\lambda_m \varrho_S}{\varrho_p^2}\left(\tanh m_--\Phi\right)\nn
    &\quad\times\bigg(1-\frac{\lambda_m^2\lambda_S}{(\lambda_m+\lambda_S)\lambda_D^2}\bigg)\bigg]\,.
\end{align}
\Cref{eq:asymptotic-RD,eq:asymptotic-RD-conditions,eq:rho-aver-explicit} form a closed system for $\overline{\rho}_\pm(t,z)$. 
However, analytical treatment is impeded by the non-linear dependence of the chemical potentials.
In the following section, the system is linearized near equilibrium.

\section{Charging dynamics near equilibrium}\label{sec:near_eq_chaging}

The pore is in equilibrium (eq) when the reaction fluxes vanish, $\mathcal{J}_\text{rct}^\text{eq}(t,z)=0$, which happens when the forward and backward reaction fluxes balance, $\mathcal{J}_{0}=\mathcal{J}_{f}^\text{eq}=\mathcal{J}_{b}^\text{eq}\geq 0$.
Note that $ec_0\mathcal{J}_{0}$ is the exchange current density \cite{bard2022electrochemical,lasia1995impedance}.
In addition, we assume that the ions in the pore are in equilibrium with the reservoir, and, because $\mu_\pm=0$ in the reservoir, so it is in the pore, $\mu^\text{eq}_\pm =0$. 
Calculating \cref{eq:phi-explicit} for $\mu^\text{eq}_\pm = 0$, we find that the centerline  electrostatic potential vanishes at equilibrium, $\phi^\text{eq}_{c}(z)=0$. 

As the chemical potential does not depend on the radial coordinate, we find the cationic concentration at the OHP as $\rho_+^\text{eq}(t,\varrho_S) = \exp\left[-\phi^\text{eq}(t,\varrho_S)\right]$.
Setting $\mathcal{J}_\text{rct}^\text{eq}=0$ in \cref{eq:FBV-def} then yields the equilibrium potential $\Phi^\text{eq} =\ln \left(k_f/k_b\right)$.
The pore can only reach or be in equilibrium when the applied potential is $\Phi=\Phi^\text{eq}$. 
When the applied potential differs from the equilibrium potential by $\delta \Phi \equiv \Phi - \Phi^\text{eq} \neq 0$, there is a nonzero reaction flux. 

\subsection{Linear reaction flux}
From hereon, we restrict our study to the \textit{linear regime} wherein the applied potential $\Phi$ is both small ($\Phi\ll1$) and close to the equilibrium potential ($|\delta \Phi| = |\Phi-\Phi^\text{eq}| \ll 1$). 
In this case, $\mathcal{J}_f/\mathcal{J}_b$ can be rewritten to
\begin{align}
\label{eq:linear-affinity}
    \frac{\mathcal{J}_f}{\mathcal{J}_b} &= \ln\left(e^{\Phi^\text{eq}}e^{\mu_+(t,z) - \phi(t,\varrho_S,z)}e^{- \Phi+\phi(t,\varrho_S,z)}\right) \nn 
    &= \delta\mu_+(t,z) - \delta\Phi\,,
\end{align}
where $\delta\mu_+(t,z) \equiv \mu_+(t,z)-\mu_\pm^\text{eq}$ and where we used that the chemical potential does not depend on the $r$-coordinate [\textit{viz.}~\cref{eq:asymptotic-equilibrium}] and $\mu_\pm^\text{eq} = 0$ implying $\delta\mu_+(t,z) = \mu_+(t,z)$. Using \cref{eq:linear-affinity}, we linearize the flux $\mathcal{J}_\text{rct} =\mathcal{J}_f -\mathcal{J}_b$ as
\begin{align}
\label{eq:flux-linear}
    \mathcal{J}_\text{rct} &= \mathcal{J}_f\left(1 - \frac{\mathcal{J}_b}{\mathcal{J}_f}\right) \approx \mathcal{J}_0\left[1-\exp\ln \left(\frac{\mathcal{J}_b}{\mathcal{J}_f}\right) \right]\nn 
    & \approx -\mathcal{J}_0\ln \frac{\mathcal{J}_b}{\mathcal{J}_f} = \mathcal{J}_{0}(\mu_+ - \delta\Phi)\,.
\end{align}
Approximating $\exp[\mathcal{O}(\Phi)] \approx 1$, we find $\mathcal{J}_{0}\approx \sqrt{k_fk_b }$.

\subsection{Linear charging dynamics}\label{sec:linearchargingdynamics}
So far, the only restriction we put on the initial ionic density was rotational symmetry, $\rho_\pm(0,z,r,\theta)=\rho_\pm(0,z,r)\equiv\rho_\pm^\text{ic}(z,r)$.
From hereon, we consider a case where $\rho_\pm^\text{ic}$ in \cref{eq:asymptotic-RD-bc-3} equals the equilibrium ion density $\rho_\pm^\text{ic} = \rho_\pm^\text{eq}$, which depends on the radial coordinate only [see \cref{eq:ic-phi}]. 
Using \cref{eq:rho-aver-explicit} for $\mu_{\pm}^\text{eq}=0$, we find $\overline{\rho}_\pm^\text{eq} = 1 \mp \Lambda\Phi^\text{eq}$; hence, $\overline{q}^\text{eq}=\overline{\rho}_+^\text{eq} - \overline{\rho}_-^\text{eq} = - 2\Phi^\text{eq}\Lambda$, with
\begin{align}
\label{eq:Lambda-def}
    \Lambda \equiv \frac{2\lambda_D^2}{(\lambda_D + \lambda_S)\varrho_p}\,,
\end{align}
which is the equilibrium version of $\Lambda_m$; in \cite{aslyamov2022analytical} we discussed how $\Lambda $ relates to the system's Dukhin number.

At equilibrium, the Poisson equation and boundary conditions read 
$-2\lambda_D^2\nabla^2\phi^\text{eq}(r)=\rho_{+}^\text{eq}(r)-\rho_{-}^\text{eq}(r)$, 
with $\phi^\text{eq}(\varrho_p)=\Phi_\text{eq}$ and $\partial_r\phi^\text{eq}(0)=0$, 
which shows that $\rho_{\pm}^\text{eq}(r)$ depend only on the radial coordinate.  
Likewise, we write the Poisson equation and boundary conditions for $t=0^{+}$ as 
$-2\lambda_D^2\nabla^2\phi^\text{ic}(r)=\rho_{+}^\text{eq}-\rho_{-}^\text{eq}$, 
with $\phi^\text{ic}(\varrho_p)=\Phi$ and $\partial_r\phi^\text{ic}(0)=0$.  
Comparing the two, we find a solution 
\begin{align}
\label{eq:ic-phi}
    \phi^\text{ic}(r)=\phi^\text{eq}(r)+\delta\Phi\,.
\end{align}

In the linear regime, since $|\rho_{\pm}(t,z,r)-\rho_\pm^\text{eq}(r)|=\mathcal{O}(\delta\Phi)$, the chemical potentials can be calculated from \cref{eq:general_model_mu} at $r=0$ as $\mu_{\pm}=\ln[\rho_\pm^\text{eq}(0)+\mathcal{O}(\delta\Phi)]\pm\phi_c$, which are of the order $\mu_{\pm} = \mathcal{O}(\Phi)$. 
This means that all expressions in \cref{sec:poisson} containing $\tanh$ or $\exp$ of $\mu_{\pm}$ or $m_{\pm}$ could have been linearized, but this was not apparent at that point.
Here, we have $\lambda_m/\lambda_D=1+\mathcal{O}(\Phi)$ and $\Lambda_m/\Lambda=1+\mathcal{O}(\Phi)$. 
Moreover, \cref{eq:rho-aver-explicit} simplifies to
\begin{equation}
\label{eq:rho-aver-linear}
    \overline{\rho}_{\pm}=1+m_+\pm\left(m_--\Phi\right)\Lambda+\mathcal{O}(\Phi^2)\,,
\end{equation}
yielding the following cross sectional-averaged charge and salt densities, 
\begin{subequations}\label{eq:q-s-linear}
\begin{align}
    \overline{q}&=\overline{\rho}_+-\overline{\rho}_-=2(m_--\Phi)\Lambda+\mathcal{O}(\Phi^2)\,,\\
    \overline{s}&=\overline{\rho}_++\overline{\rho}_-=2+2m_++\mathcal{O}(\Phi^2)\,.
\end{align}
\end{subequations}
Using $\overline{\rho}_\pm = 1 +\mathcal{O}(\Phi)$ [\cref{eq:rho-aver-linear}], we linearize \cref{eq:asymptotic-RD} as
\begin{subequations}\label{eq:linear-RD}
\begin{align}
    \partial_{t} \overline{\rho}_{+}&=D\partial_{z}^2\mu_+-2\mathcal{J}_{0}(\mu_+ - \delta\Phi)\,,\label{eq:linear-RD-plus}\\
    \partial_{t} \overline{\rho}_{-}&= D \partial_{z}^2\mu_-\,,\label{eq:linear-RD-minus}
\end{align}
\end{subequations}
which we rewrite in terms of $m_{\pm}$ using \cref{eq:q-s-linear} and $\delta\mu_+ = \mu_+ = m_+ + m_-$, 
\begin{subequations}\label{eq:linear-RD-m}
\begin{align}
    \partial_{t}m_+&=D\partial_{z}^2m_+-\mathcal{J}_{0}\left(m_++m_- - \delta\Phi\right)\,,\label{eq:linear-RD-m-plus}\\
    \Lambda\partial_{t}m_-&=D\partial_{z}^2m_--\mathcal{J}_{0}\left(m_++m_- - \delta\Phi\right)\,.\label{eq:linear-RD-m-minus}
\end{align}
\end{subequations}
Linearizing \cref{eq:phi-c-tanh_m}, we find $\phi_c(t,z)= m_-(t,z)$, which means that \cref{eq:linear-RD-m-minus} describes the evolution of the centerline potential.

The differential equations \eqref{eq:linear-RD-m} come with initial and boundary conditions for $m_{\pm}(t,z)$,
\begin{subequations}\label{eq:linear-beyond-TLM}
\begin{align}
    \partial_z m_{\pm}(t,\ell_p) & = 0\,,\label{eq:linear-beyond-TLM-z-1}\\
    D\partial_z m_\pm(t, 0) &= \mathcal{L}m_\pm(t,0)\,,\label{eq:linear-beyond-TLM-z-2}\\
    m_-(0^+,z) & = \delta\Phi\,, \label{eq:linear-beyond-TLM-ic-1}\\
    m_+(0^+,z) & = 0 \,. \label{eq:linear-beyond-TLM-ic-2}
\end{align}
\end{subequations}
Here, to derive the boundary conditions in \cref{eq:linear-beyond-TLM-z-2} we used \cref{eq:asymptotic-RD-bc-1} for small applied potentials as $D\overline{\rho}_\pm\partial_z \mu_\pm = D\partial_z \mu_\pm + \mathcal{O}(\Phi^2)$.
To derive the initial conditions in  \cref{eq:linear-beyond-TLM-ic-1,eq:linear-beyond-TLM-ic-2}, we use \cref{eq:ic-phi} for the initial conditions in terms of the chemical potentials as
\begin{align}
    \mu_\pm(0^+,z) = \pm [\phi^\text{eq}(0)+\delta\Phi] = \pm \delta\Phi\,,
\end{align}
which results in \cref{eq:linear-beyond-TLM-ic-1,eq:linear-beyond-TLM-ic-2}.

In this article, we will focus on the early-time [\cref{sec:faradiacTLmodel}] and steady-state [\cref{sec:steady_state}] behavior as predicted by \cref{eq:linear-RD-m,eq:linear-beyond-TLM}.
In doing so, we will ignore salt and charge transport that sets up at intermediate times, the study of which we leave for future work.
With a macrohomogeneous electrode model,  Devan and coworkers \cite{devan2004analytical} found that the slow-moving salt concentration causes a second ``diffusion'' arc in the Nyquist plot of the impedance.

\section{Early times: Faradaic TL model}\label{sec:faradiacTLmodel}

We now solve \cref{eq:linear-RD-m} analytically for early times. To do so, we first show that \cref{eq:linear-RD-m} is equivalent to the dynamics of the centerline potential [\cref{eq:phi_c_eq+bcs}]. Next, we demonstrate that this dynamics can be mapped onto an $RC$ circuit, whose elements are determined by the electrolyte and pore-surface properties [\cref{TLCircuitanalysis,RC-parameters}]. Finally, we present analytical solutions for both the centerline dynamics and the spatiotemporal potential distributions [\cref{sec:TL-solution}]. 

\subsection{Time scale separation}

\Cref{eq:linear-RD-m} shows that $m_+(t,z)$ and $m_-(t,z)$ have significantly different time scales due to the parameter $\Lambda$. 
As $\Lambda\ll1$, $m_-(t,z)$ relaxes much faster than $m_+(t,z)$.
Hence, for $t\lessapprox \Lambda \ell_p^2/D$, $m_-(t,z)$ evolves while $m_+(t,z)$ remains constant,
\begin{equation}
\label{eq:symmetric-cp}
    m_+(t,z) = \text{const} = 0\,,    
\end{equation}
where the constant is zero as, right after ($t=0^+$) applying the potential $\Phi$, the concentration has not yet changed, but the chemical potentials are $\mu_{\pm}(0^+,z)=\pm\Phi$.

Using \cref{eq:symmetric-cp,eq:phi-c-tanh_m}, we find
\begin{equation}
    \phi_c(t,z)= m_-(t,z) = \mu_+(t,z) \,,  \label{eq:phi_c-mu}
\end{equation}
and inserting it into \cref{eq:linear-RD-m-minus}, we write
\begin{subequations}\label{eq:phi_c_eq+bcs}
\begin{align}
    \frac{\Lambda}{D}\partial_{t}\phi_c &=\partial_{z}^2\phi_c -\frac{\mathcal{J}_{0}}{D}\left(\phi_c - \delta\Phi\right)\,,\label{eq:rd-phi_c}\\
\intertext{subject to conditions from \cref{eq:linear-beyond-TLM}}
    \phi_c(0,z)&=\delta\Phi\,, \label{eq:TL-ic0}\\
    \partial_z\phi_c(t,\ell_p) &= 0\,. \label{eq:TL-bc-right-def}\\
        \label{eq:TL-bc-left-def}
    \partial_{z}\phi_c(t,0)&=\frac{\lambda_D^2}{A_p\varepsilon D}\frac{1}{R_{r}}\phi_c(t,0)\,.
\end{align}    
\end{subequations}
Here, for \cref{eq:TL-bc-left-def}, we used the connection of the boundary conditions in \cref{eq:linear-beyond-TLM-z-2} and Ohm's law. 
On the one hand, the ionic current $\overline{I}(t,0) \equiv e c_0 A_p [\overline{j}_+(t,0)-\overline{j}_-(t,0)]$ through a cross section at $z=0$ can be expressed in terms of the ionic fluxes as 
\begin{equation}
\label{eq:bc-linear-regime-3}
    \overline{I}(t,0) = e c_0 A_p \mathcal{L} (\mu_+ - \mu_-) = 2 e c_0 A_p  \mathcal{L} \phi_c(t,0)\,, 
\end{equation}
On the other hand, assuming the potential drop in the reservoir to be $\phi_c$, we apply Ohm law's as
\begin{align}
    \overline{I}(t,0)&=\frac{k_B T \phi_c(t,0)}{e\,R_r}\,,\label{eq:bc-linear-regime-1}
\end{align}
with $R_r$ the resistance of the reservoir \cite{yang2022simulating}. Comparing these expressions, we find the Onsager coefficient as
\begin{align}
\label{eq:Onsager-R_r}
    \mathcal{L} = \frac{\lambda_D^2}{\varepsilon A_p}\frac{1}{R_r}\,,
\end{align}
which, inserted into \cref{eq:linear-beyond-TLM-z-2} yields \cref{eq:TL-bc-left-def}.

Starting from the full PNP-FNP equations to study a pore of large aspect ratio  ($\ell_p\gg \varrho_p$), with a thin EDL ($\varrho_p\gg \lambda_D$) and Stern layer ($\varrho_p\gg \lambda_S$), and subject to a potential close to the equilibrium potential $\Phi^\text{eq}$, we have arrived at \cref{eq:phi_c_eq+bcs}---the first of two main result of the paper.
In the next section, we show that \cref{eq:phi_c_eq+bcs} also governs the potential on the top horizontal line in the Faradaic TL circuit in \cref{fig:fig-1}(c).

\subsection{TL circuit analysis}\label{TLCircuitanalysis}
From hereon, we switch back to the dimensional electrostatic potential $\psi$ and applied potential $\Psi(t)=\psi(t,\varrho_p,z)$. 
We consider again the pore as defined in \cref{sec:setup}, but now discuss its equivalent circuit shown in \cref{fig:fig-1}(c).
The pore has a total capacitance $C$, an electrolyte resistance $R_p$, and a Faradaic transfer resistance $R_F$.
But because these resistances and capacitance are distributed over the pore, it does not charge as a circuit connection of the elements $R_p, R_F$, and $C$. 
Instead, we represent the pore through a TL circuit containing $n$ identical modules, each with elements with resistances $r_p = R_p /n$ and $r_F = R_F n$, and capacitance $c=C/n$. 
We partition the $z$-coordinate as $z=k\,dz$ for $k=0,1,\dots, n$ with the step $dz$ such that $n\, dz = \ell_p$. 
The TL then contains $n+1$ points with potentials $\psi_{c,k} = \psi_c(t, k dz)$, and resistances and capacitances
\begin{align}\label{eq:circuit_parameters}
    r_p = R_p \frac{dz}{\ell_p}\,,\quad c=C \frac{dz}{\ell_p}\,\quad r_F = R_F\frac{\ell_p}{dz}\,.
\end{align}

We consider three subsequent elementary modules none of which at the circuit's start or end. 
Kirchhoff's junction rule relates the currents $I_{p,k}$ and $I_{p,k+1}$ through the elementary resistances $r_p$ of two subsequent modules $k$ and $k+1$ (see \cref{fig:fig-1}c),
\begin{align}
\label{eq:Kirch}
    I_{p,k} = I_{p,k+1} + I_{F,k} - \frac{d q_{c,k}}{dt}\,,
\end{align}
where $I_{F,k}$ and $q_{c,k}$ are the current through the Faradaic resistance and charge on the capacitor in the $k$th module. 
Using Ohm's law we find the difference of pore currents $ I_{p,k+1} - I_{p,k}$,
\begin{align}
    \label{eq:I_k}
    r_p I_{p,k} &= \psi_{k} - \psi_{c,k-1}\,,\nn
    r_p (I_{p,k+1} - I_{p,k}) &= \psi_{c,k+1} - 2\psi_{c,k} +\psi_{c,k-1}\,.
\end{align}
The capacitors in \cref{fig:fig-1}(c) account for the EDL and Stern layer capacitances, accumulating the charge 
\begin{align}
    q_{c,k} = - c(\Psi - \psi_{c,k})\,.\label{eq:qc_k}
\end{align}
Kirchhoff's law relates the potential difference between the circuit's top and bottom horizontal wires to the potential drop across Faradaic resistance and the bias voltage, 
\begin{align}
    r_F I_{F,k} + \Psi^\text{eq} = \Psi - \psi_{c,k} \,.\label{eq:IF_k}
\end{align}
Combining \cref{eq:Kirch,eq:I_k,eq:qc_k,eq:IF_k} we arrive at
$r_p c d_t \psi_{c,k} =  \psi_{c,k+1} - 2 \psi_{c,k} + \psi_{c,k-1} + (r_p/r_F)( \Psi - \Psi^\text{eq} - \psi_{c,k})$, which, with \cref{eq:circuit_parameters} and $\psi_{c,k} = \psi_c(t, k\, dz)$ reads
\begin{align}\label{eq:TL-pde-discrete}
   R_p C d_t \psi_c = \ell_p^2\frac{\psi_{c,k+1} - 2 \psi_{c,k} + \psi_{c,k-1}}{dz^2}  + \frac{R_p}{R_F}( \Psi - \Psi^\text{eq} - \psi_c)\,.
\end{align}

At the left boundary ($z=0$), current conservation through the resistance $R_r$ and the elementary resistance of the first modules implies $\psi_{c,0}/R_r = (\psi_{c,1} - \psi_{c,0})/r_p$, or, using \cref{eq:circuit_parameters},
\begin{align}\label{eq:TLcircuit_left}
   \frac{R_p}{R_r} \psi_{c,0} = \frac{\psi_{c,1} - \psi_{c,0}}{ dz}\ell_p\,.
\end{align}

For the right boundary, we note that $I_{n+1} = 0$. 
Using \cref{eq:I_k,eq:qc_k,eq:IF_k} for the last module, we arrive at
\begin{align}\label{eq:TLcircuit_right}
    R_p C d_t \psi_{c,n} = - \ell_p^2\frac{\psi_{c,n} - \psi_{c,n-1}}{dz^2}  + \frac{R_p}{R_F}( \Psi - \Psi^\text{eq} - \psi_{c,n})\,.
\end{align}

In the continuum limit of an infinite number of modules $n\to \infty$ (and $dz\to 0$), we find
\begin{subequations}\label{eq:TL-system}
\begin{align}
     R_p C  \partial_{t}\psi_c&=\ell_p^2 \partial_{z}^{2}\psi_c+\frac{R_p}{R_F}(\delta\Psi-\psi_c)\,,\label{eq:TL-pde}\\
    \psi_c(0,z)&=\delta\Psi\,, \label{eq:TL-ic}\\
    \ell_p\partial_{z}\psi_c(t,0)&=\frac{R_p}{R_{r}}\psi_c(t,0), \label{eq:TL-bc-left}\\
    \partial_{z}\psi_c(t,\ell_p)&=0\,,\label{eq:TL-bc-right}
\end{align}
\end{subequations}
where \cref{eq:TL-pde-discrete} turned to \cref{eq:TL-pde}, \cref{eq:TL-ic0} turned to \cref{eq:TL-ic}, \cref{eq:TLcircuit_left} turned to \cref{eq:TL-bc-left}, and \cref{eq:TLcircuit_right} turned to $\psi_n = \psi_{n-1}$; hence, to \cref{eq:TL-bc-right}. 
Finally, we notice that for the variables $\psi_c^{DL} = \Psi \psi_c/\delta\Psi$, the system \cref{eq:TL-pde} becomes the well-known Faradaic TL model of de Levie \cite{levie1967electrochemical}; see Eq. (43) in \cite{pedersen2023equivalent}. 

\Cref{eq:TL-system} generalizes previous Faradaic TL equation: when $\Psi^\text{eq}=0$, we have $\delta\Psi = \Psi$ and \cref{eq:TL-pde} coincides with Eq.~(34) in \cite{pedersen2023equivalent} or Eq.~(92) in \cite{levie1967electrochemical}, which describes only the reactions with $k_{f}=k_{b}$.

\subsection{Circuit parameters}\label{RC-parameters}
To make the connection between \cref{eq:phi_c_eq+bcs,eq:TL-system} explicit, we express the pore's capacitance $C$ and its electrolyte and Faradaic resistances $R_p$ and $R_F$ in terms of electrolyte and pore-surface properties.

From the Nernst-Planck equation follows a dilute electrolyte's resistivity, $\rho=\lambda_D^2/(\varepsilon D)$, so the electrolyte resistance of the cylindrical pore is
\begin{equation}
   R_p =\frac{\lambda_D^2 \ell_p}{\varepsilon D A_p}\,.\label{eq:R_p}
\end{equation}
The pore's capacitance $C$ is given by a harmonic mean of the Stern layer $C_{S}$ and double layer $C_\text{EDL}$ capacitances, $1/C=1/C_{S}+1/C_\text{EDL}$.
For small applied potentials, the EDL and Stern layer can be treated as coaxial cylindrical dielectric capacitors whose electrode separation are set by the Debye and Stern lengths, respectively.
A coaxial cylindrical capacitor with radii $\varrho_1$ and $\varrho_2$ ($\varrho_2>\varrho_1$) and length $\ell_p$ has a capacitance $C_\text{cyl}=2\pi \varepsilon \ell_p/\ln (\varrho_2/\varrho_1)$. 
By inserting $\varrho_1=\varrho_p-\lambda_S$ and $\varrho_2=\varrho_p$ for $C_{S}$ and $\varrho_1=\varrho_p-\lambda_S-\lambda_D$ and $\varrho_2=\varrho_p-\lambda_S$ for $C_\text{EDL}$, and taking $\varrho_p\gg \lambda_D, \lambda_S$, we find
\begin{equation}\label{eq:C}
    C\approx\frac{2\pi \varepsilon \varrho_p\ell_p}{\lambda_S+\lambda_D}\,.
\end{equation}

Last, using \cref{eq:IF_k} we write  
\begin{equation}
    I_{F,k} = \frac{\psi_c - \delta\Psi}{R_F}\,\frac{dz}{\ell_p}\,.\label{eq:I_F-1}
\end{equation}
Microscopically, the current $I_{F,k}$ is caused by electrons released or consumed in the reaction \eqref{eq:redox}. 
Therefore we can write
\begin{align}
\label{eq:I_F-2}
     I_{F,k} &= e c_0 \int_{0}^{\varrho_p}\int_{0}^{2\pi} \mathcal{J}_\text{rct}\delta\Big(\frac{r}{\varrho_S}-1\Big)\, rdr\, d\theta dz\,,\nn
     &=\frac{2e^2c_0}{k_BT}\mathcal{J}_{0}(\psi_c - \delta\Psi)\,A_pdz\,.
\end{align}
Combining \cref{eq:I_F-1,eq:I_F-2} gives $R_F = k_B T/(2c_0e^2\ell_pA_p\mathcal{J}_{0})$, equivalent to Eq.~(12) (for $\alpha=1/2$) as stated by Lasia \cite{lasia1995impedance}\footnote{The equivalence follows if one calculates the characteristic current of electrons produced or consumed in \cref{eq:redox} as $2\pi \ell_p\int e c_0 \mathcal{J}_{0}\delta(r/\varrho_p-1) r dr =2\ell_pA_p\mathcal{J}_{0}$ or as $2\pi\varrho_p\ell_p e j_0$, where $j_0$ is the surface density kinetics $[j_0]=1/(m^2 sec)$. 
Lasia used the latter version and his $j_0=\mathcal{J}\varrho_p$.}.
We rewrite $R_F$ using \cref{eq:R_p} to
\begin{equation}
\label{eq:R_F}
R_F=\frac{\lambda_D^2}{\varepsilon \ell_p A_p}\frac{1}{\mathcal{J}_{0}}=\frac{\lambda_D^2}{\varepsilon \ell_p A_p}\frac{1}{\sqrt{k_f k_b}}=\frac{DR_p}{\ell_p^2\mathcal{J}_{0}}\,.
\end{equation}

Inserting \cref{eq:C,eq:R_p,eq:R_F} into \cref{eq:TL-system}, we recover \cref{eq:phi_c_eq+bcs}.
We have thus shown that the FBV-PNP equations predict a pore's charging to be captured---in the linear regime, for long pores, dilute electrolytes, and thin EDLs---by a Faradaic TL circuit whose lumped parameters coincide with ad hoc estimates thereof.
In doing so, we have extended our previous work \cite{aslyamov2022analytical}, wherein we showed such an equivalence for the charging of a blocking pore.

\subsection{Faradaic TL equation \eqref{eq:TL-system} solution}
\label{sec:TL-solution}

\subsubsection{Overpotential representation}
We rewrite \cref{eq:TL-system} in terms of the overpotential $\eta \equiv \Phi - \phi_c - \Phi_{\text{eq}}$,
\begin{subequations}\label{eq:TL-system-eta}
\begin{align}
    R_p C \partial_{t}\eta &= \ell_p^2\partial_{z}^{2}\eta - \text{Da}\,\, \eta\,,\label{eq:TL-pde-eta}\\
    \eta(0,z)&= 0\,, \label{eq:TL-ic-eta}\\
    \ell_p\partial_{z}\eta(t,0)&=\text{Bi}[\eta(t,0)-\delta\Psi]\,,\label{eq:TL-bc-left-eta}\\
    \partial_{z}\eta(t,\ell_p)&=0\,,\label{eq:TL-bc-right-eta}
\end{align}
\end{subequations}
Here, we adopted the notation of \cite{biesheuvel2011diffuse}, where the Damk\"{o}lher number $\text{Da} = R_p/R_F$ compares the contributions to charging of charge-transfer and migration, while the Biot number $\text{Bi}=R_p/R_r$ compares the rate of migration in the bulk reservoir to that within the pore. 
We notice that \cref{eq:Onsager-R_r} implies $\text{Bi} = \ell_p\mathcal{L}/D$. 

\Cref{eq:TL-system-eta} has the same form as Eqs.~(25) and (31) of Ref.~\cite{biesheuvel2011diffuse}. 
With this observation, we can use their Eq.~(32) to solve \cref{eq:TL-system-eta} (for a full derivation, see \cref{sec:appendix-TL-solution})
\begin{align}
\label{eq:eta-sol}
    &\frac{\eta(t,z)}{\delta\Psi}= \frac{\cosh\left[\sqrt{\text{Da}}(z/\ell_p-1)\right]}{\text{Bi}^{-1}\sqrt{\text{Da}}\sinh \sqrt{\text{Da}}+ \cosh \sqrt{\text{Da}}}\\
    &-\sum_{j\ge1}\frac{4\beta_j^2}{\beta_j^2+\text{Da}}\frac{\sin \beta_j\cos\left[\beta_j(z/\ell_p-1)\right]e^{ -(\beta_j^2+\text{Da})t/(R_pC)}}{2\beta_j+\sin 2\beta_j}\nonumber
\end{align}
with $\beta_j$ the solutions of $\beta_j\tan \beta_j=\text{Bi}$. 

Even though Ref.~\cite{biesheuvel2011diffuse} solved exactly the same TL model problem [Eqs.~(25) and (31) there, \cref{eq:TL-system-eta} here], their model is different from ours as their equations contain a rescaled time and a different Damk\"{o}hler number.
In their derivation, they do not linearize $\text{Da}$ for small $\Phi$, but even if one linearizes their expression thus, their resulting Damk\"{o}hler number depends explicitly on $\lambda_S/\lambda_D$ (ours does not), which can be traced back to their macrohomogeneous setup [Eq.~(5) and (6) there].
In their case,  $\text{Da} = \mathcal{O}(\lambda_D/\varrho_p)$, which implies that Faradaic currents are small and do not significantly affect the diffusive propagation of the overpotential in the equivalent $RC$-transmission line.
In our model, $\text{Da}>1$ is possible as well.

\begin{figure}
    \centering
    \includegraphics[width=\linewidth]{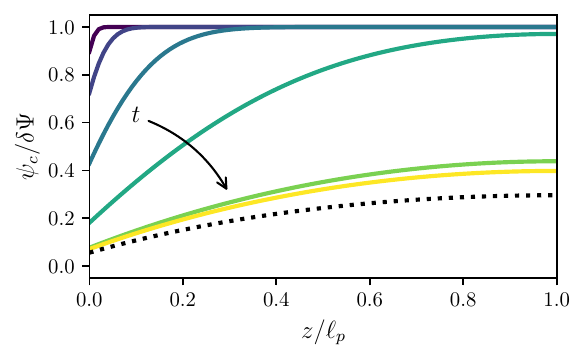}
    \caption{The centerline potential ${\psi}_c(t,z)$ \cref{eq:phi_c-sol} for  $\text{Bi}=R_p/R_r=10$, and $\text{Da}=R_p/R_F=1$. The curves correspond to $t/(R_p C)=10^{-4}, 10^{-3}, 10^{-2}, 10^{-1}, 1, 10$ (purple to yellow). The black dotted line shows the steady state value $\psi_c^\text{ss}(z)$ [\cref{eq:psi_c-ss}].}
    \label{fig:phi_d}
\end{figure}

\subsubsection{Centerline dynamics}

Using \cref{eq:eta-sol}, we find the centerline potential
\begin{align}
\label{eq:phi_c-sol}
    &\frac{\psi_c(t,z)}{\delta\Psi}=1 -  \frac{\cosh\left[\sqrt{\text{Da}}(z/\ell_p-1)\right]}{\text{Bi}^{-1}\sqrt{\text{Da}}\sinh \sqrt{\text{Da}}+ \cosh \sqrt{\text{Da}}}\\
    &+\sum_{j\ge1}\frac{4\beta_j^2}{\beta_j^2+\text{Da}}\frac{\sin \beta_j\cos\left[\beta_j(z/\ell_p-1)\right]e^{ -(\beta_j^2+\text{Da})t/(R_pC)}}{2\beta_j+\sin 2\beta_j}\nonumber\,.
\end{align}

At late times, $\psi_c(t,z)$ relaxes with the TL timescale $\tau_{TL}=R_pC/(\beta_1^2+\text{Da})$, where $\beta_1=\min(\beta_j)$. 
\Cref{fig:phi_d} shows \cref{eq:phi_c-sol} for $\text{Bi}=10$ and $\text{Da}=1$. 
In absence of Faradaic reactions, $\text{Da}\,\to0$, we have $\delta\Psi=\Psi$. 
In that case, \cref{eq:phi_c-sol} reduces to Posey and Morozumi's Eq.~(5a) in \cite{janssen2021transmission}.

\subsubsection{Spatiotemporal potential $\psi(t,r,z)$}
The pore's center-line potential [\cref{eq:phi_c-sol}] gives access to the time-dependent chemical potential $\mu_\pm(t,z)=\pm\phi_c(t,z)$ [\cref{eq:phi_c-mu}], which, by \cref{eq:phi-explicit}, gives access to the potential $\psi(t,r,z)=\phi(t,r,z)kT/e $ in the entire pore. 
In the linear regime,
\begin{align}
\label{eq:phi-distr}
   \psi(t,r,z) = \psi_c(t,z) - \frac{\lambda_D}{\lambda_D+\lambda_S}\left[\psi_c(t,z) -\Psi \right] \frac{{\rm I}_0(r/\lambda_D)}{{\rm I}_0(\varrho_S/\lambda_D)}\,.
\end{align}
The above equation correctly reduces to $ \psi(t,0,z) = \psi_c(t,z)$ on the pore's centerline, as the last term drops in our case of interest,  $\varrho_S\gg\lambda_D$.

\begin{figure}
    \centering
    \includegraphics[width=\linewidth]{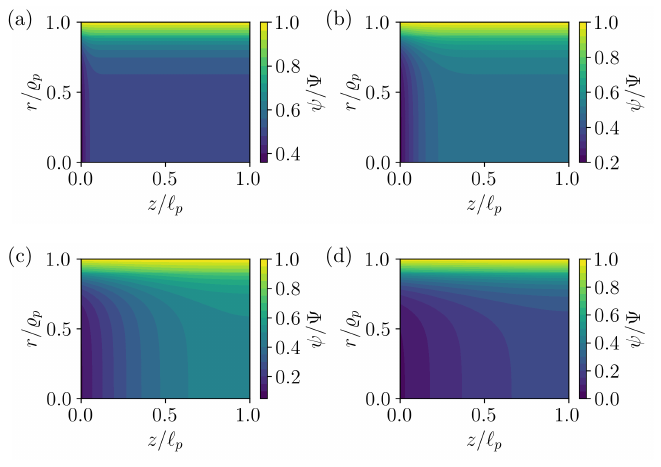}
    \caption{The spatial distribution of the potential $\psi(t,r,z)$ at times $t/(R_p C)=0.001,0.01,0.1,1$ from (a) to (d), respectively. The heatmap is calculated by \cref{eq:phi-distr} for  $\text{Bi}=R_p/R_r=10$, $\text{Da}=R_p/R_F=1$, 
    $\lambda_D=\lambda_S=0.1 \varrho_p$,
    $\Lambda = R_p C D/\ell_p^2=0.1$, 
    $e\Psi/(kT) = - 0.2$, and
    $e\Psi^\text{eq}/(kT) = - 0.1$.
    }
    \label{fig:fig-3}
\end{figure}

\Cref{fig:fig-3} shows the analytical spatiotemporal potential $\psi(t,r,z)$ as determined by \cref{eq:phi-distr,eq:phi_c-sol} for several times. 
The heatmap shows that, especially at earlier times, the spatial distribution is strikingly different from the center-line curve.

\section{Steady state}\label{sec:steady_state}

The previous section dealt with early time charging dynamics, valid for $t<\Lambda \ell_p^2/D$. 
Hence, the result \cref{eq:phi_c-sol} obtained cannot be used to determine the system's steady state ($t\to\infty$). 
Here, we solve \cref{eq:linear-RD} in the steady state, where the left-hand sides of the equations are zero. 
For the non-reacting anions,
\begin{subequations}
    \begin{align}
        0 &= \ell_p^2 \partial_{z}^2\mu_-^\text{ss}\,,\\
        D\partial_z\mu^\text{ss}_-(z=0)&=\mathcal{L}\mu^\text{ss}_-(z=0)\,,\\
        \partial_z\mu^\text{ss}_-(z=\ell_p) &= 0\,,
    \end{align}
\end{subequations}
which has a trivial solution $\mu^\text{ss}_- = 0$. 
Hence, the nonreacting anions reach equilibrium at steady state; see similar argumentation in IIIC of \cite{biesheuvel2011diffuse}. 
Next, we can write $\mu^\text{ss}_+ = 2m^\text{ss}_\pm = 2e\psi^\text{ss}_c/(k_BT)$ which, inserted into the steady-state form of \cref{eq:linear-RD}, gives
\begin{subequations}
    \begin{align}
        0 &= \ell_p^2 \partial_{z}^2\psi^\text{ss}_c + \text{Da} (2\psi^\text{ss}_c-\delta\Psi)\,,\\
        \ell_p\partial_z\psi_c^\text{ss}(z=0)&=\text{Bi}\,\psi_c(0)\,,\\
        \partial_z\psi_c^\text{ss}(z=\ell_p) &= 0\,,
    \end{align}
\end{subequations}
which we solve as
\begin{align}
\label{eq:psi_c-ss}
    \psi_c^\text{ss} = \frac{\delta\Psi}{2} - \frac{\delta\Psi}{2} \frac{\cosh\left[\sqrt{2\text{Da}}(z/\ell_p-1)\right]}{\text{Bi}^{-1}\sqrt{2\text{Da}}\sinh \sqrt{2\text{Da}}+ \cosh \sqrt{2\text{Da}}}\,,
\end{align}
and which is equivalent to Eq.~(42) of \cite{biesheuvel2011diffuse}.

\Cref{fig:phi_d}(b) shows ${\psi}_c^\text{ss}(t,z)$ [\cref{eq:psi_c-ss}] for $\text{Bi}=5$ and several $\text{Da}$.
We see that as $\text{Da}$ increases, the difference between the entrance and end centerline potentials become stronger. 
This plot is similar to Fig.~3 in Lasia \cite{lasia1995impedance}, where the overpotential is shown for several exchange current densities---varying $\mathcal{J}_{0}$, we vary $R_F$, hence, the Damk\"{o}hler number.
As Lasia, we find that the potential drop between the pore's surface and its centerline is largest near the pore entrance, so charge transfer will happen primarily in the pore mouth region. 
The expressions derived by him and plotted in Fig.~3 there, however, do not quantitatively agree with ours.

\begin{figure}
    \centering
        \includegraphics[width=\linewidth]{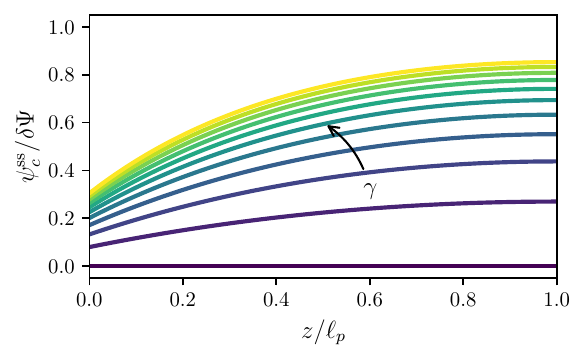}
    \caption{The scaled steady-state centerline potential $\psi_c^\text{ss}/\delta\Psi$ [\cref{eq:psi_c-ss}] for $\text{Bi}= 5$ and $\text{Da}$ from $0$ to $2.5$ with the step $0.25$ (solid curves from purple to yellow).}
    \label{fig:phi_d}
\end{figure}

\section{Potential of zero-charge}\label{sec:PZC}
A key characteristic of electrochemical systems is their \textit{potential of the zero charge (PZC)}---the potential $\Psi^\text{pzc}$ for which  the system is uncharged \cite{trasatti1999potential,huang2020obtaining}. 
The PZC can be controlled by using electrolytes with ions of different sizes \cite{fedorov2008ionic} and chemically active electrodes \cite{cohen2011enhanced}. 
The latter dependency can be used to boost the effectiveness of deionization devices: more salt can be adsorbed by a device using a positive electrode with a negatively shifted PZC and a negative electrode with a positively shifted PZC \cite{porada2013review}. 

To find the $\Psi^\text{pzc}$ of our pore, we first calculate the steady state current $I^\text{ss}$ measured at the entrance to the pore
\begin{align}
    I^\text{ss} = \frac{\ell_p\partial_z\psi^\text{ss}_c(0)}{R_p} = \frac{\delta\Psi}{2R_p} \frac{\sqrt{2\text{Da}}\sinh\sqrt{2\text{Da}}}{\text{Bi}^{-1}\sqrt{2\text{Da}}\sinh \sqrt{2\text{Da}}+ \cosh \sqrt{2\text{Da}}}\,,
\end{align}
which allows us to rewrite \cref{eq:psi_c-ss} as
\begin{align}
\label{eq:psi_c-ss-Z}
    \psi_c^\text{ss} = \frac{\delta\Psi}{2} - \frac{\delta\Psi R_p}{\hat{Z}(0)} \frac{\cosh\left[\sqrt{2\text{Da}}(z/\ell_p-1)\right]}{\sqrt{2\text{Da}}\sinh\sqrt{2\text{Da}}}\,,
\end{align}
where we introduced the zero-frequency impedance from its step response to a small potential deviation (see details in \cite{pedersen2023equivalent}) as
\begin{align}
    \hat{Z}(0) \equiv \frac{\delta\Psi}{I^\text{ss}} = R_r + R_p\frac{\coth \sqrt{2\text{Da}}}{\sqrt{2\text{Da}}}\,.
\end{align}

\begin{figure}
    \centering
    \includegraphics[width=\linewidth]{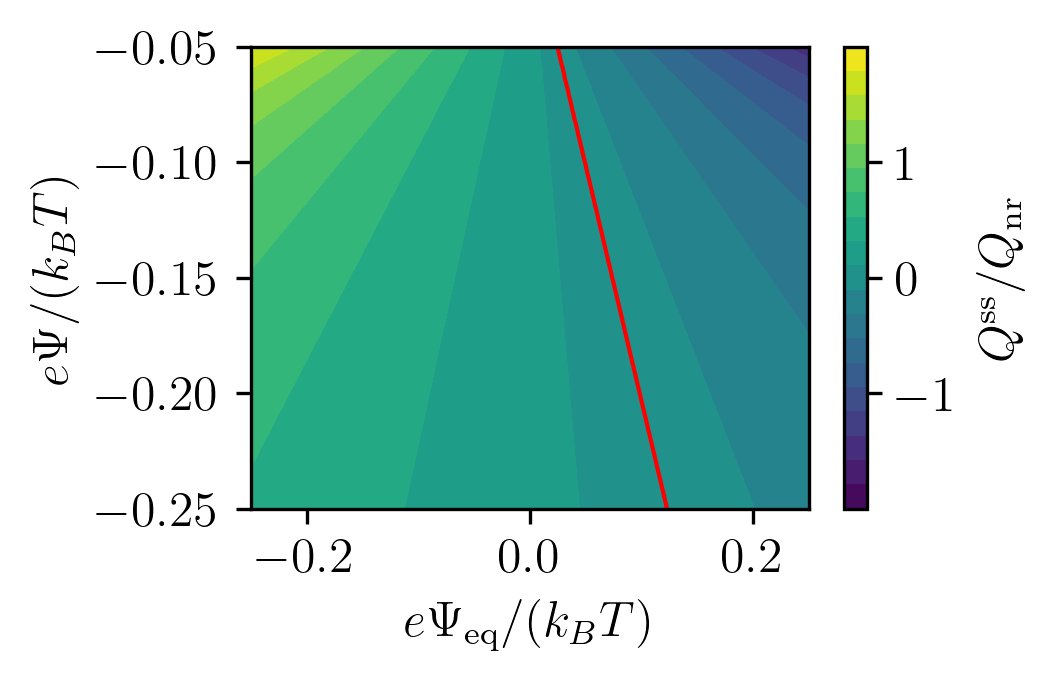}
    \caption{The steady-state ratio $Q^\text{ss}/Q^\text{nr}$ calculated in the plane of $\Psi_\text{eq}$ and $\Psi$ for $\text{Bi}=1$ and $\text{Da}=2$. 
    The red line lies at $Q^\text{ss} = 0$ and illustrates \cref{eq:PZC}.
    }
    \label{fig:fig-2-Qss}
\end{figure}

Now, we calculate the steady-state charge by integrating the charge density \cref{eq:q-s-linear} over the pore length, 
\begin{align}
\label{eq:charge-ss}
   Q^\text{ss}&=e c_0 A_p\int_0^{\ell_p}\overline{q}^\text{ss}(z)\,dz = \frac{e^2 c_0 A_p \Lambda}{k_B T}\int_0^{\ell_p} [\psi^\text{ss}_c(z)-\Psi]\,dz\nonumber\\
    &=\frac{\delta\Psi R_p +\Psi_\text{eq}\text{Da}\,\hat{Z}(0)}{2\text{Da}\,\hat{Z}(0)\Psi}Q^\text{nr}\,,
\end{align}
where $Q^\text{nr}=-C \Psi$ is the charge of a non-reacting pore ($\text{Da}=0$) with $C=2e^2 c_0 A_p\ell_p \Lambda/(k_B T)$.
\Cref{fig:fig-2-Qss} shows $Q^\text{ss}/Q^\text{nr}$ [\cref{eq:charge-ss}] for various $\Psi^\text{eq}$ and $R_F$.
The figure shows how chemical reactions affect the charge on the pore's surface, displaying a region of the enhanced capacity with $Q^\text{ss}>Q^\text{nr}>0$ with the same applied potential.

We now find the pore's PZC (the potential for which it is uncharged), by setting  $Q^\text{ss}=0$ in \cref{eq:charge-ss}.
This gives
\begin{align}
\label{eq:PZC}
    \Psi^\text{pzc} = \Psi^\text{eq}\left(1 - \frac{\hat{Z}(0)}{R_F}\right)\,,
\end{align}
which is the second of the two main results of this paper.

Unlike prior porous electrode models \cite{paasch1993theory,devan2004analytical,biesheuvel2011diffuse}, our model has $\Psi^\text{eq} \neq 0$.
Consequently, at the PZC, part of the pore contains a positive charge, balanced by another part with a negative charge. 
\Cref{eq:PZC} includes influences from the electrode material through $\Psi_\text{eq}$ \cite{trasatti1999potential}, the porous geometry through $\hat{Z}(0)$, and charge‐transfer kinetics through $R_F\propto 1/\mathcal{J}_0$ \cite{li2022surface}. 

In reverse, one can use \cref{eq:PZC} to find $R_F$ as
\begin{align}
\label{eq:R_F-exper}
R_F = -\frac{\Psi^\text{eq}}{\Psi^\text{pzc}-\Psi^\text{eq}}\hat{Z}(0) = -\frac{\Psi^\text{eq}}{I^\text{pzc}_\text{ss}}\,,
\end{align}
where we used $I_\text{ss} = \lim_{s\to 0 }s\hat{I}(s)$ and where $I_\text{pzc}^\text{ss}$ is the current in $\Psi^\text{pzc}$.
Thus, \cref{eq:R_F-exper} can be used to experimentally determine the Faradaic resistance for systems with known PZC.

\section{Discussion}\label{sec:discussion}

Our analysis has several features present in the macrohomogenous models of Paasch \cite{paasch1993theory}, Devan \cite{devan2004analytical}, Biesheuvel \cite{biesheuvel2011diffuse}, and their respective coworkers.
As in \cite{devan2004analytical}, for instance, we start from a dynamical equation for charge carriers and the FBV equation (they use the BV equation), and after linearizing for small potentials, arrive at two coupled PDEs for the charge density and overpotential [Eqs. (17) and (18) in \cite{devan2004analytical}, \cref{eq:linear-RD-m} here].
Comparing to \cite{biesheuvel2011diffuse}, we use the same Frumkin correction and our models have the same equilibrium potentials. Moreover, in \cref{sec:biesheuvel_limits} we show that our $\mathcal{J}_\text{rct}$ simplifies to identical expressions as found in \cite{biesheuvel2011diffuse} in cases where EDLs are either thin or thick compared to the Stern layer.
Some notable differences between our work and these macrohomogenous electrode models are the following.
(i) The macrohomogenous models do not give access to the spatiotemporal potential and charge profiles inside individual pores. 
Our 3d PNP approach does yield these profiles [\emph{viz}.~\cref{fig:fig-3}], allowing us to track EDL formation in the radial direction.
Frumkin's correction, used before in \cite{biesheuvel2011diffuse}, is based on the geometric insight that charge transfer happens primarily at the OHP---in our 3d PNP approach this correction is included fully explicitly.
The detailed spatiotemporal insight into a pore's charging offered by the PNP approach will be especially relevant if ionic potentials can be tracked in pores, for instance, using XPS \cite{kutbay}
(ii)~Ref.~\cite{biesheuvel2011diffuse} assumes micropores to be in quasi-equilibrium with macropores;  no such assumption enters our model.
(iii)~The circuits drawn by Paasch, Devan, Biesheuvel and their respective coworkers do not contain bias voltage sources.
In fact, at an intermediate step [\cref{eq:TLmodelredefined}] of our derivation, we solved a reaction-diffusion problem for the overpotential $\eta$, whose circuit representation contains no bias voltage sources either. 
In our view, the circuit in \cref{fig:fig-1}(c) gives a better physical representation of our system as, in an experiment, as one controls the electrode potential $\Psi$ and not the overpotential $\eta$.

\section{Conclusions}\label{sec:conclusions}
We studied the charging of a long cylindrical pore through Faradaic reactions and the formation of thin EDLs, in response to a small applied potential, close to the equilibrium potential of the system.
The first main result of this paper is that, under these conditions, the PNP-FBV equations can be solved in terms of the potential at the centerline of the pore, which is governed by the reaction-diffusion equation \eqref{eq:phi_c_eq+bcs}.
We show that \cref{eq:phi_c_eq+bcs} also governs the behavior of the TL circuit in \cref{fig:fig-1}(c).
The circuit in \cref{fig:fig-1}(c) is similar to prior TL models accounting for Faradaic reactions at a pore's surface \cite{levie1967electrochemical,pedersen2023equivalent},  except that it contains a voltage source of potential $\Psi^\text{eq}$ in every branch of the circuit, biasing the Faradaic reactions.
Such bias sources also appear in the Hodgkin--Huxley model \cite{hodgkin1952quantitative}, where $\Psi^\text{eq}$ corresponds to ``reversal potential'' of the ion channel stalling the current in the ionic channels \cite{dayan2005theoretical}.

The second main result of this paper is our analytical expression of the PZC [\cref{eq:PZC}], which suggests a new way to measure it. 
However, we derived \cref{eq:PZC} assuming the pore to be long.
Future work may study the generality of our results for shorter pores or overlapping EDLs, and study the intermediate-time behavior of \cref{eq:linear-beyond-TLM}.
Other direction of potential interest are to consider electrolytes with unequal diffusivities, to extend our work to multistep and multielectron reactions, consider reaction kinetics beyond the FBV model~\cite{bazant2023unified}, or to consider reactions where the reaction product stays in solution.
To better model real porous electrodes, one could account for electrode resistivity \cite{paasch1993theory,devan2004analytical} or combine pores into a network~\cite{biesheuvel2012electrochemistry,henrique2024network}.
Last, one could incorporate molecular roughness of electrode surfaces~\cite{Aslyamov2022Properties} through modified Poisson equations~\cite{aslyamov2021electrolyte}, which influence the local cross-sectional equilibrium.

\section{Acknowledgments}
T.A. and M.E. acknowledge the financial support from, respectively, 
project ThermoElectroChem (C23/MS/18060819) from Fonds National de la Recherche-FNR, Luxembourg;
project TheCirco (INTER/FNRS/20/15074473) funded by FRS-FNRS (Belgium) and FNR (Luxembourg).
M.J. was supported by a FRIPRO grant from The Research Council of Norway (Project No. 345079).

\appendix

\section{Derivation of \cref{eq:eta-sol}}\label{sec:appendix-TL-solution}
Using the dimensionless parameters $\bar{t}=t/R_p C$ and $\bar{z}=z/\ell_p$, we rewrite \cref{eq:TL-system-eta} to
\begin{subequations}
\label{eq:TLmodelredefined}
\begin{align}
    \partial_{t}\eta_{}&=\partial_{z}^{2}\eta_{}-\text{Da}\,\, \eta_{}\,,\\
    \eta_{}(0,z)&=0\,, \\
    \partial_{z}\eta_{}(t,0)&=\text{Bi}\left[\eta_{}(t,0)-\delta\Psi\right], \\
    \partial_{z}\eta_{}(t,1)&=0\,,
\end{align}
\end{subequations}
where we dropped the bars.
We perform Laplace transformations on the above equation, where we write $\hat{\eta}(s)\equiv \mathscr{L}\left\{\eta(t)\right\}\equiv\int_{0}^{\infty} \eta(t)\exp{(-ts)}\,dt$ and use 
$\mathscr{L}\left\{\partial_t \eta(t,z)\right\}=s\hat{\eta}(s,z)-\eta(0,z)$.
Writing $\varsigma^2\equiv\text{Da}+s$, we find
\begin{subequations}
\begin{align}
    \varsigma^2 \hat{\eta}_{}&=\partial_{z}^{2}\hat{\eta}_{}\label{eq:1a} \\
    \partial_{z}\hat{\eta}_{}(s,0)&=\text{Bi}\left(\hat{\eta}_{}(s,0)-\frac{\delta\Psi}{s}\right),\label{eq:1b} \\
    \partial_{z}\hat{\eta}_{}(s,1)&=0\,,\label{eq:1c}
\end{align}
\end{subequations}
which is solved by 
\begin{equation}\label{eq:phi_d_intermediate}
    \hat{\eta}_{}(s,z)=\frac{\delta\Psi}{s}\frac{\cosh\left[\varsigma(z-1)\right]}{\text{Bi}^{-1}\varsigma\sinh \varsigma+ \cosh \varsigma}\,.
\end{equation}

Determining $\eta_{}(t,z)=\mathcal{L}^{-1}\left\{\hat{\eta}_{}(s,z)\right\}$ requires performing an inverse Laplace transformation. 
By the residue theorem, $\eta_{}(t,z)=\sum_{s\in s_{\ell}}\text{Res}\bm{\big(}\hat{\eta}_{}(s,z)\exp(st),s_{\ell}\bm{\big)}$, where the poles $s_{\ell}=\{s_{0}, s^{\star}_{j} \}$ of $\hat{f}(x, s)$ are located at $s_{0}=0$, and $s^{\star}_{j}$ which solve
\begin{equation}\label{eq:transcendental_eq-0}
    \text{Bi}^{-1}\varsigma_j\sinh \varsigma_j+ \cosh \varsigma_j=0\,,
\end{equation}
where $\varsigma_j^2=\text{Da}+s^{\star}_{j}$.

The pole $s_{0}=0$ gives the steady-state solution,
\begin{equation}\label{eq:steadystate}
    \eta_{}^\text{ss}=\delta\Psi \frac{\cosh\left[\sqrt{\text{Da}}(z-1)\right]}{\text{Bi}^{-1}\sqrt{\text{Da}}\sinh \sqrt{\text{Da}}+ \cosh \sqrt{\text{Da}}}\,.
\end{equation}

For the poles at $s^{\star}_{j}$, we expand 
\begin{align}
    &\text{Bi}^{-1}\varsigma\sinh \varsigma+ \cosh \varsigma\overset{s\to s^{\star}_{j}}{=}\nn
    &\quad= \frac{\partial (\text{Bi}^{-1}\varsigma\sinh \varsigma+ \cosh \varsigma)}{\partial s}\bigg|_{s=s^{\star}_{j}}\left(s- s^{\star}_{j}\right)\nn
    &\quad=\frac{1}{2}\left(\frac{1}{\varsigma_j}(\text{Bi}^{-1}+1)\sinh \varsigma_j+ \text{Bi}^{-1}\cosh \varsigma_j\right)\left(s- s^{\star}_{j}\right)\,.
\end{align}
Therefore, we find
\begin{align}
    &\eta(t,z)-\eta_{}^\text{ss}(z)=\sum_{j\ge1}\text{Res}\bm{\big(}\hat{\eta}_{}\exp(st),s^{\star}_{j}\bm{\big)}\\
    &=\sum_{j\ge1}\frac{\delta\Psi}{s^{\star}_j}\frac{2\cosh\left[\varsigma_j(z-1)\right]}{\varsigma_j^{-1}(\text{Bi}^{-1}+1)\sinh \varsigma_j+ \text{Bi}^{-1}\cosh \varsigma_j}\exp( s_j^{\star} t)
    \,.\nonumber
\end{align}
Writing $\varsigma_j=i\beta_j$, we find $s^{\star}_{j}=-\beta_j^2-\text{Da}$, so
\begin{align}
\label{eq:denominator}
    &\eta(t,z)-\eta_{}^\text{ss}(z)=\\
    &=-\sum_{j\ge1}\frac{2\delta \Psi }{ \beta_j^2+\text{Da}}\frac{\beta_j\cos\left[\beta_j(z-1)\right] \exp\left[-(\beta_j^2+\text{Da}) t\right]}{(\text{Bi}^{-1}+1)\sin \beta_j+ \text{Bi}^{-1}\beta_j\cos \beta_j}\,.\nonumber
\end{align}
Multiplying \cref{eq:denominator} by $\sin \beta_j/\sin \beta_j$, we find a denominator we that we rewrite using $\beta_j\tan \beta_j=\text{Bi}$ [\cref{eq:transcendental_eq-0}] twice: 
$(\text{Bi}^{-1}+1)\sin^2 \beta_j+ \text{Bi}^{-1}\beta_j\sin \beta_j\cos \beta_j =\sin 2\beta_j/(2\beta_j)+1$.
We find
\begin{align}
    &\eta(t,z)-\eta_{}^\text{ss}(z)=\\
    &=-\sum_{j\ge1}\frac{4\delta \Psi\beta_j^2}{ \beta_j^2+\text{Da}}\frac{\sin \beta_j\cos\left[\beta_j(z-1)\right]}{2\beta_j+\sin 2\beta_j}e^{ \displaystyle-(\beta_j^2+\text{Da}) t}\,,\nonumber
\end{align}
which, with \cref{eq:steadystate} yields \cref{eq:eta-sol}.
We checked \cref{eq:eta-sol} against a numerical inverse Laplace transform of $\hat{\eta}_{}(s,z)$ [\cref{eq:phi_d_intermediate}] and found identical results.

\section{Special limits of reaction kinetics}\label{sec:biesheuvel_limits}
Typically, the Stern length is around the size of the ion diameters.
Conversely, the Debye length varies over multiple decades depending on the salt concentration.  
For dense electrolytes such as ionic liquids, the Debye length is approximately $0.1$ times the molecular diameter; see \cite{bazant2011double}). 
Hence, the ratio $\lambda_S/\lambda_D$ can vary much and, following Biesheuvel and coworkers \cite{biesheuvel2011diffuse}, we now consider special forms of FBV kinetics in the Gouy--Chapman limit ($\lambda_S/\lambda_D\to0$) and Helmholtz limit ($\lambda_S/\lambda_D\to\infty$).

In the Gouy--Chapman limit, $\phi(t,\varrho_S,z)=\Phi$ and the FBV kinetics [\cref{eq:FBV-def}] takes the form 
\begin{align}
\label{eq:FBV-GC}
    \mathcal{J}_\text{rct}=k_f\rho_{+}(t,\varrho_S,z)-k_b = k_b\big(e^{- \eta} - 1\big)\,,
\end{align}
where $\eta = \Phi - \phi_c-\Phi_\text{eq}$ is the overpotential and
where we used $\ln \rho_{+}(t,\varrho_S,z) = \mu_+ - \Phi$ with $\mu_+=\phi_c$ and $\Phi^\text{eq}=\ln \left(k_f/k_b\right)$. 
Our \cref{eq:FBV-GC} is equivalent to the upper expression in Eq.~(16) of \cite{biesheuvel2011diffuse}. 

In the Helmholtz limit, using $\lambda_m \approx \lambda_D \ll \lambda_S$ in \cref{eq:phi-explicit}, we find $\phi_S \approx \phi_c$. 
Thus, the FBV current takes the form $\mathcal{J}_\text{rct} = k_f \rho_+(t,\varrho_S,z) \exp[-(\Phi - \phi_c)/2 ] - k_b \exp[(\Phi - \phi_c)/2] $, in agreement with Lasia's Eq.~(12) \cite{lasia1995impedance}. 
We notice that $\rho_+(t,\varrho_S,z) = \exp(\mu_+ - \phi_c) = 1$, where we used \cref{eq:general_model_mu} for $\phi(t,\varrho_S,z)=\phi_c$. 
Then we have 
\begin{align}
\label{eq:FBV-H}
    \mathcal{J}_\text{rct} = 2\sqrt{k_fk_b}\sinh(\eta)\,,
\end{align}
which is equivalent to lower expression in Eq.~(16) of \cite{biesheuvel2011diffuse}. 

\bibliographystyle{apsrev4-2}   
\bibliography{biblio}
\end{document}